\documentstyle[12pt]{article}

\def\be{\begin{equation}}       
\def\ee{\end{equation}}         

\def\bea{\begin{eqnarray}}      
\def\eea{\end{eqnarray}}
                                
\def\ba{\begin{array}}
\def\ea{\end{array}}

\def\eq{\begin{equation}}
\def\eqe{\end{equation}}
\def\eqa{\begin{eqnarray}}
\def\eqae{\end{eqnarray}}
\def\ena{\end{eqnarray}}

\def\eg{{\it e.g.~}}
\def\ie{{\it i.e.~}}

\def\Tr{{\rm Tr}}

\def\unit{1 \hskip-.3em \raise2pt\hbox{$ \scriptstyle |$ } }

\def\a{\alpha}
\def\b{\beta}
\def\c{\gamma} 
\def\d{\delta}
\def\e{\epsilon}           
\def\f{\phi}               
   
\def\g{\gamma}

\def\k{\kappa}                    
\def\l{\lambda}
\def\m{\mu}
\def\n{\nu}
  
\def\p{\pi}                
\def\r{\rho}                                     
\def\s{\sigma}                                   

\def\x{\xi}

\def\D{\Delta}
\def\F{\Phi}
\def\G{\Gamma}

\def\P{\Pi}

\def\ca{{\cal A}}

\def\cf{{\cal F}}

\def\ck{{\cal K}}
\def\cl{{\cal L}}

\def\cn{{\cal N}}
\def\co{{\cal O}}

\def\half{{1 \over 2}}

\def\bop#1{\setbox0=\hbox{$#1M$}\mkern1.5mu
        \vbox{\hrule height0pt depth.04\ht0
        \hbox{\vrule width.04\ht0 height.9\ht0 \kern.9\ht0
        \vrule width.04\ht0}\hrule height.04\ht0}\mkern1.5mu}
\def\Box{{\mathpalette\bop{}}}                        
\def\pa{\partial}                              
\def\de{\nabla}                                       

\def\>{\rangle} 

\def\<{\langle} 
\def\Dsl{D \hskip-.6em \raise1pt\hbox{$ / $ } }

\def\leftrightarrowfill{$\mathsurround=0pt \mathord\leftarrow 
\mkern-6mu
       \cleaders\hbox{$\mkern-2mu \mathord- \mkern-2mu$}\hfill
       \mkern-6mu \mathord\rightarrow$}
\def\dvec#1{\vbox{\ialign{##\crcr
       \leftrightarrowfill\crcr\noalign{\kern-1pt\nointerlineskip}
       $\hfil\displaystyle{#1}\hfil$\crcr}}}          
\def\hook#1{{\vrule height#1pt width0.4pt depth0pt}}
\def\leftrighthookfill#1{$\mathsurround=0pt \mathord\hook#1
       \hrulefill\mathord\hook#1$}
\def\underhook#1{\vtop{\ialign{##\crcr                 
       $\hfil\displaystyle{#1}\hfil$\crcr
       \noalign{\kern-1pt\nointerlineskip\vskip2pt}
       \leftrighthookfill5\crcr}}}
\def\smallunderhook#1{\vtop{\ialign{##\crcr      
       $\hfil\scriptstyle{#1}\hfil$\crcr
       \noalign{\kern-1pt\nointerlineskip\vskip2pt}
       \leftrighthookfill3\crcr}}}

\def\to{\rightarrow}
\def\tf{\tilde{\f}}

\def\pa{\partial}

\def\+{\oplus}

\def\de{\mbox{d}} 

\def\nonu{\nonumber \\{}}
\def\half{{1 \over 2}}
\def\Tr{{\rm Tr}\, }
\def\tB{\tilde{B}}

\begin{document}
\begin{flushright}
MIT-CTP-3166, \ PUTP-1999\\
DAMTP-2001-63, \ ROM2F/2001/30\\
{\tt hep-th/0112119}
\end{flushright}

\begin{center}
{\Large\bf
Holographic Renormalization}
\vskip .6truecm

{\large\bf Massimo Bianchi,${}^{\star}${}\footnote{
{\tt Massimo.Bianchi@roma2.infn.it}} 
 Daniel Z. Freedman${}^{\dagger}${}\footnote{
{\tt dzf@math.mit.edu}} \\
and Kostas Skenderis${}^{\ddagger}$\footnote{{\tt
kostas@feynman.princeton.edu}}}\\
\vskip 0.3truecm
${}^{\star}$ {\it 
Dipartimento di Fisica, Universit\`a di Roma ``Tor Vergata'' \\
00133 Rome, Italy}
\vskip 0.3truemm
${}^{\dagger}$ {\it Department of Mathematics 
and Center for Theoretical Physics, MIT \\
Cambridge MA 02139, USA }
\vskip 0.3truemm
${}^{\ddagger}$ {\it Physics Department, Princeton University \\
Princeton, NJ 08544, USA}

\end{center}
\vskip .3truecm
\begin{abstract}
We systematically develop the procedure of holographic renormalization
for RG flows dual to asymptotically AdS domain walls. All divergences
of the on-shell bulk action can be cancelled by adding covariant local
boundary counterterms determined by the near-boundary behavior
of bulk fields. This procedure defines a renormalized action from
which correlation functions are obtained by functional 
differentiation.
The correlators are finite and well behaved at coincident points. 
Ward identities, corrected for anomalies, are satisfied. The
correlators depend on parts of the solution of the bulk field 
equations
which are not determined by near-boundary analysis. In principle a
full nonlinear solution is required, but one can solve linearized
fluctuation equations to define a bulk-to-boundary propagator from
which 2-point correlation functions are easily obtained. We carry out
the procedure explicitly for two known RG flows obtained from the
maximal gauged D=5 supergravity theory, obtaining new results on
correlators of vector currents and related scalar operators and giving
further details on a recent analysis of the stress tensor sector.

\end{abstract}

\newpage

\section{Introduction}

The AdS/CFT correspondence \cite{Malda,Wit,Gubs} has provided a way 
to 
obtain strong coupling results on conformal field theories from string
theory and supergravity. The techniques can be extended to describe
non-conformal theories which are deformations of $CFT's$ either by
addition of relevant operators to the Lagrangian or by turning on
$VEV's$ for these operators. The gravity duals of such theories are
domain wall solutions of $(d+1)$-dimensional bulk theories whose 
isometry group is
the $d$-dimensional Poincar\'e group. Many such classical solutions 
are
known \cite{FGPW,FGPW2,bs,GPPZ,pstarinets,pz0002172,warner,review}. 
Given such a solution one may address the problem
of calculating correlation functions of operators in the deformed 
$CFT.$
Much information about 2-point correlation functions has been 
obtained
using a method which was formulated quite early in the game 
\cite{Gubs,
fmmr}. But, as we discuss below, this method does not deal adequately
with divergences and it is not straightforward to obtain
correlators of the energy-momentum tensor and vector currents. The
purpose of this paper is to apply the method of holographic 
renormalization
\cite{HS,KSS} to this problem. A finite renormalized action $S_{{\rm 
ren}}$ 
can be
constructed systematically using this method. $S_{{\rm ren}}$ is a
functional of boundary data for the bulk fields which are the sources 
of
operators in the field theory. Renormalized correlation functions are 
then 
obtained by functional differentiation, and these correlators satisfy
appropriate Ward identities including anomalies. The calculation of 
1-
point functions is particularly clear in this method.  

The domain wall spacetimes asymptotically approach $AdS_{d+1}$ geometries 
and the divergences which occur in the holographic computation of
correlation functions are due to the infinite volume of these 
geometries.
These $IR$ divergences correspond to the $UV$ divergences which occur 
in
field theory when composite operators approach coincident points. 

We now outline how the $IR$ divergences are resolved in the method of
holographic renormalization. The first step is to express bulk fields
as series expansions in the radial coordinate which is transverse to
the boundary. This allows one to determine an asymptotic solution of 
the 
field equations given arbitrary Dirichlet boundary conditions. The
solution is obtained by substituting the series expansions in the
nonlinear bulk field equations and solving term-by-term. This process
is called near-boundary analysis. Roughly speaking it determines the
first $\D - d/2$ terms in the expansion where $\D$ is the dimension 
of the
operator dual to a given bulk field. Beyond this point additional
information is required to determine a unique solution, as expected 
since
we have second order field equations and only the
Dirichlet condition at the $AdS$ boundary has been 
specified so far. To specify a
unique solution we require that fluctuations about the domain wall
background vanish in the deep interior. The coefficients determined 
by 
near-boundary analysis are local functions of the boundary data, while
higher order terms can be non-local.

The next step of the method is to construct $S_{{\rm ren}}$ by a 
process of
regularization and renormalization. Both the concepts and details of 
this
construction are important. The bulk theory is regulated by
cutting off the spacetime at a large finite value of the radial
coordinate. The series solution is then inserted in the regulated 
classical
action. One then observes that the on-shell action contains a finite
number of terms which diverge as the cutoff is removed. These 
divergences
involve only coefficients in the solution which are fixed by 
near-boundary
analysis. They can be cancelled by adding counterterms to the
action which are expressed as invariant local functionals of the 
induced metric and other fields at the boundary of the cutoff 
space-time.
These fields depend locally on the Dirichlet data at the true 
boundary which
is approached as the cutoff is removed.
The entire process is similar to what is done in field theory, and
there is an ambiguity of finite local counterterms which is just the
usual scheme dependence in field theory. The sum of the regulated 
action
plus counterterms is finite as the cutoff is removed, and $S_{{\rm 
ren}}$ is
defined by this limit. 
By virtue of its construction $S_{{\rm ren}}$ is invariant under 5D
diffeomorphisms, except for the ones which generates Weyl
transformations of the boundary metric. The violation of Weyl 
invariance
is manifested in logarithmically divergent counterterms, and conformal
anomalies can be read directly from these terms.

Finite correlation functions can then be obtained by functional 
differentiation of $S_{{\rm ren}}$ with respect to sources.
This computation, which involves a limiting procedure described in 
Sec 4 
below, can be carried out in all generality, and the
correlation functions then involve the lowest order
series coefficients which are not determined by near-boundary 
analysis. 
This is to be expected since correlation functions are non-local.
By contrast,
field theory UV divergences and anomalies are local, and as such 
one can obtain them by the near-boundary analysis. This also 
implies that Ward identities can be established by the
near-boundary analysis. 

The domain-wall spacetimes that describe the RG-flows generically 
contain
naked singularities in the interior. Near the singularities the 
curvatures are large and the supergravity approximation breaks 
down. A full string theory treatment is presumably necessary to
resolve the singularity. However, at 
least for quantities that do not depend crucially on IR properties,
one should still be able to use supergravity. Taking this point of 
view, a very important issue is what boundary conditions are imposed
at the singularity. In  principle this information should come from 
string theory.
Here we take as a working assumption that the fluctuations vanish 
at the singularity. 

Ideally one would like to have an exact solution of the nonlinear 
field
equations with arbitrary Dirichlet boundary conditions and suitable
behavior in the interior. The coefficients in the boundary expansion 
not
determined by local analysis could then be extracted and $n$-point
correlation functions computed. This problem is now intractable. 
However,
it is not difficult to solve the fluctuation equations linearized 
about
the background. The solutions are the extension to $RG$-flows of 
bulk-to-boundary propagators in the pure $AdS$ background \cite{Wit},
and 2-point correlators are easily obtained from their boundary 
asymptotics.
Higher point 
functions can be obtained by means of Witten diagrams, but we shall 
not 
pursue this here.

In the conventional method \cite{Gubs,fmmr} for calculating 2-point
functions, one also imposes a cutoff at large radius and solves the
linearized fluctuation equation with Dirichlet boundary condition at
the cutoff (as opposed to the scaled Dirichlet problem on the full
spacetime in holographic renormalization). The second variation of
the cutoff on-shell action is then computed in momentum space 
yielding an expression
containing singular powers of the cutoff times integer powers of $p$
plus nonsingular terms in the cutoff which are non-analytic in $p$.
The 2-point function is defined as the leading non-analytic term.
The polynomial terms in $p$ which are dropped are just contact
$\d$-function terms in the position space correlator, which are
scheme dependent in field theory and largely unphysical. The 
non-analytic
term has an absorptive part in $p$ which correctly gives the 2-point
function for separated points in $x$-space. Although this method is
quite efficient, it is not fully satisfactory since it is not correct
in general simply to drop divergent terms in correlation functions.
Subtractions in different correlators should be mutually consistent,
and this is guaranteed if the subtractions are made by local
covariant counterterms. Further, to our knowledge, there is no 
systematic
method other than holographic renormalization to obtain correct 
finite values
for 1-point functions. We hope to convince readers that holographic
renormalization is a formalism which resolves divergences in all 
correlators and makes Ward identities and their anomalies manifest
{\it ab initio}. Further, the procedure need only by carried out once 
for
each given bulk action with results applying to all classical 
solutions. 
Considerable insight into the relation of holography and field theory
is also exhibited, including the fact that the cancellation of $UV$
divergences does not depend on the $IR$ physics.

In previous work \cite{howtogo} we have applied holographic 
renormalization
to the computation of 2-point functions of the stress tensor and the
operator dual to the scalar field in the domain wall background for
two specific $RG$-flows, the GPPZ flow \cite{GPPZ} describing $\cn 
=4$ SYM
theory perturbed by an $\cn =1$ supersymmetric mass term, and a 
Coulomb
branch flow \cite{FGPW2,bs} describing a disc distribution of 
$D3$-branes.
Essentially equivalent results were simultaneously presented in 
\cite{mueck}
with earlier progress in \cite{theisen}.
In this paper we present details of the near 
boundary analysis of the field equations which were omitted in 
\cite{howtogo},
and we extend the method to the computation of correlators of some
vector currents. The emphasis here is on longitudinal components of 
the
currents, since the transverse 2-point functions can be obtained 
\cite{Anatomy,bs2} by more conventional methods. However, 
we do
summarize recent developments on the issue of Goldstone poles
in transverse correlators.

In Sections 2 and 3 we review the background solutions for the $GPPZ$ 
and
$CB$ flows and the Lagrangians and equations of motion required to 
obtain
the correlation functions we study. In Section 4 we describe the 
procedure
of near-boundary analysis and how it leads to 1-point functions (with 
sources)
that satisfy correct anomalous Ward identities and from which higher 
point
correlators can be obtained. In Section 5 details of the near 
boundary 
analysis of the stress tensor sector of both flows are given. These 
results
were already used in \cite{howtogo} to obtain correlation functions. 
Sections
6,7,8 are devoted to the current sector. Near boundary analysis for 
vector
fields, linear fluctuations, and correlation functions are all 
discussed.
In the Appendix a group theory lemma on the preserved flavor symmetry 
on
the Coulomb branch of $\cn=4$ SYM theory is stated and proved.


\section{Holographic RG-flows}
\setcounter{equation}{0}

In this section we review the method generally used to obtain 
supersymmetric domain wall background solutions of 5-dimensional 
supergravity theories. The first step is to apply symmetry arguments
\cite{warner} to obtain a consistent truncation of the full bulk
theory, usually $D=5, \cn = 8$ gauged supergravity, to a small
number of scalar fields interacting with gravity. In the cases we
discuss a single scalar is sufficient and the truncated 
supergravity action is\footnote{
Our curvature conventions are as follows
$R_{\mu\nu\kappa}{}^\lambda=\pa_\mu \G_{\nu\kappa}{}^\lambda + 
\G_{\mu\rho}{}^\lambda \G_{\nu\kappa}{}^\rho - \mu
\leftrightarrow \nu$ and $R_{\mu\nu}=R_{\mu\lambda\nu}{}^\lambda$. 
They differ 
by an overall sign from the conventions in \cite{Notes,Anatomy}.}
\be
S_0 = {N^2 \over 2 \p^2} \int_M \de^5x \sqrt{G} \left[{1 \over 4} R  
+\half G^{\m \n} \pa_\m \F \pa_\n \F + V(\F) \right]
-\half \int_{\pa M} \sqrt{\c} \ck
\label{scaction}
\ee
where $\ck$ is the trace of the second fundamental form.
We work in Euclidean signature. We assume that the potential
$V(\F)$ has a stationary point at $\F=0$. By a constant 
Weyl transformation \cite{BST} one may set $V(\F{=}0)=-3$, so that
the action admits a pure $AdS$ solution with scale $L=1$ in string
units.  As discussed in \cite{howtogo}, 
the Weyl rescaling produces the overall factor of $N^2/2\p^2$ in 
(\ref{scaction}), and 
the holographically computed
correlation functions agree with those of the undeformed
$\cn =4$ SYM at short distance. 

The second step is to look for solutions of the Euler-Lagrange
equations with $d=4$ Poincar\'e symmetry. The most general form is
\be \label{dmw}
ds^2 = e^{2 A(r)} \delta_{ij} dx^i dx^j + dr^2  \qquad , \qquad
\Phi = \Phi(r)
\ee
Equivalent forms involving a different radial variable will be used
later. With radial variable $r$ the asymptotic boundary $AdS$ region
is $r \rightarrow \infty$, and the scale factor is exponential in
this region, i.e. $\exp (2A(r)) \rightarrow \exp (2r)$. The 
asymptotic 
behavior of the scalar field partly determined by the mass or 
equivalently 
the dimension of the operator dual to it, the two being related by 
$\Delta = 2 + \sqrt{4 + m^2}$, distinguishes solutions which are dual 
to flows
governed by operator deformations from those triggered by VEV's. 

Even within the ansatz (\ref{dmw}) it is generally difficult to solve 
the
second order field equations of (\ref{scaction}). The following 
simplifying
procedure is thus widely used. Namely if the potential $V(\F)$ is 
derivable
from a superpotential $W(\F)$,
\be \label{potential}
V={1 \over 2} 
\left[{\pa W\over \pa \F}\right]^2 - {4 \over 3} W^2,
\ee
then any solution of the first order equations 
\be \label{1storder}
{dA\over dr} =- {2 \over 3} W \qquad , \qquad {d \F\over dr} = {\pa W\over 
\pa\F}
\ee
provides a domain wall
solution for the action (\ref{scaction}) \cite{ST,DFGK}. 
These equations have been
obtained from several standpoints, namely gravitational stability
\cite{Townsend,ST},
fermion transformation rules in the (truncated) 
supergravity theory \cite{FGPW},
and the Hamilton-Jacobi framework \cite{dBVV}. 
We will be applying (\ref{potential}),
(\ref{1storder}) in a supersymmetric context associated with Killing
spinors, and the superpotential $W$ will have a critical point at
$\F =0$.

We now review the application of this formalism to the two domain 
wall 
spacetimes for which we exemplify the formalism of holographic
renormalization. It should be emphasized that the first order 
equations
apply only to the domain wall background. The full second order 
equations of motion must be studied, via holographic renormalization,
to obtain correlation functions of 
operators dual to fluctuations about the background. We
begin this analysis in Section 4.

\subsection{Coulomb branch flow} \label{cbdisc}

The first case we consider is the CB flow with $n=2$, in the notation 
of 
\cite{FGPW2}. The behavior of the scalar field 
near the boundary (see below) shows that this solution corresponds to 
a VEV of a scalar operator of dimension 2. 
The VEV breaks R-symmetry from $SO(6)$ to $SO(2) \times SO(4)$. 
The scalar that gets a VEV is the neutral  singlet 
component of Tr~($X^i X^j)$ 
in the decomposition ${\bf 20}' \rightarrow
({\bf 1}, {\bf 1})_0 \+ ({\bf 1}, {\bf 1})_{+2} \+ ({\bf 1}, {\bf 
1})_{-2} \+ 
({\bf 2}, {\bf 2})_{+1} \+ ({\bf 2}, {\bf 2})_{-1} \+ ({\bf 3}, {\bf 
3})_0 $ 
under the $SO(4) \times SO(2)$ subgroup of $SO(6)$.
Similarly, the vector currents decompose according to ${\bf 15}
\rightarrow
({\bf 1}, {\bf 1})_0 \+ ({\bf 1}, {\bf 3})_{0} \+ ({\bf 3}, {\bf 
1})_{0} \+ 
({\bf 2}, {\bf 2})_{+1} \+ ({\bf 2}, {\bf 2})_{-1}$. 
Since the system describes a distribution of $N$ D3 branes on a disk,
the $SO(2)$ symmetry is broken for any finite $N$. The corresponding 
field theory statement is that the Coulomb branch of $\cn = 4$ SYM
with any finite dimensional gauge group $G$ should not
contain any configurations where the flavor group $SO(6)$ 
is broken to a non-trivial direct product 
$SO(n)\times SO(6-n)$ with $2 \le n \le 5$. We show 
in the appendix that this is indeed the case.

As mentioned above, the flow describes the theory
at a point on the Coulomb branch of $\cn = 4$ SYM.
The superpotential  is given by 
\be
W(\F)=-e^{-{2 \F \over \sqrt{6}}} - \half e^{{4 \F \over \sqrt{6}}}.
\ee
Using (\ref{potential}), one can easily compute the potential. Near 
$\F=0$ 
it has the expansion 
\be \label{CBpot}
V(\F) = -3 - 2 \F^2 + {4 \over 3 \sqrt{6}} \F^3 + \co(\F^4)
\ee 
which exhibits a tachyonic mass $m^2 = -4$, so that $\F$ is dual to
an operator of scale dimension $\D =2$.

The domain-wall solution is expressed implicitly in terms of a 
variable
$v$. The solution and the relation between $v$ and $r$ are
\be
v=e^{\sqrt{6} \F}, \qquad e^{2A}=\ell^2{v^{2/3} \over 1-v}, \qquad 
{d v \over d r} = 2 v^{2/3} (1-v) \nonu
\ee
The boundary is at $v=1$ and the solution has a curvature 
singularity at $v=0$. The parameter $\ell$ is the radius of the disk 
of
branes \cite{FGPW2,bs} in the 10-dimensional ``lift'' of this 
solution.

The scalar field vanishes at the boundary at
the rate
\be
\F \approx -{1 \over \sqrt6} (1-v) = - {1 \over \sqrt6} \exp(-2r)
\ee
For an operator of dimension $\D =2$, the dual bulk scalar approaches 
the 
boundary either at the rate $r\exp(-2r)$ associated with an operator
deformation of the Lagrangian or at the rate $\exp(-2r)$ associated 
with
a VEV. We see that the VEV rate applies in this flow.

\subsection{GPPZ flow}

The second example we will deal with is the GPPZ flow \cite{GPPZ}.
This solution corresponds to adding an operator of dimension 3 
to the Lagrangian that gives a common mass to the 
three ${\cal N} =1$ chiral multiplets appearing in the decomposition
of the ${\cal N}=4$ vector multiplet. The solution was proposed as the
dual of pure ${\cal N}= 1$ SYM  theory. Although it does not capture 
all of the expected features of the field theory, we will use it to
illustrate the application of holographic renormalization to an
operator deformation of a $CFT_4$.

The superpotential is given by
\be
W(\F)=-{3 \over 4} \left[1+\cosh \left({2 \F \over \sqrt{3}} 
\right)\right]
\ee
Near $\F=0$, the potential has an expansion,
\be \label{gppzpot}
V(\F) = - 3 -{3 \over 2} \F^2 - {1 \over 3} \F^4 + \co(\F^6)
\ee
The mass is $m^2=-3$ indicating that the dual scalar has dimension
$\Delta = 3$. The domain-wall solution is given by
\be
\F = {\sqrt{3}\over 2} \log {1 +\sqrt{1-u} \over 1-\sqrt{1-u}}
\qquad e^{2 A} = {u \over 1-u},
\ee
where $u = 1 - \exp(-2r)$.
The boundary is at $u=1$ and the solution is singular at $u=0$.
Since $\F \approx \sqrt{3} \exp(-r)$ near the boundary, 
we are dealing with an operator deformation by a dimension 3 operator,
namely the top component of the superpotential 
$\Delta W = \sum_{I=1}^3 \Phi_I^2$.


\section{Lagrangian and Field Equations}
\setcounter{equation}{0}

We will be interested in computing correlation functions
in theories holographically described by 
domain-wall backgrounds. 
This means that we need to solve the coupled system of 
equations that involves the corresponding bulk fields and then 
evaluate the on-shell action.
For correlation functions that involve the stress tensor,
certain R-symmetry currents and scalar operators, the
bulk theory is that of gravity coupled to gauge and scalar
fields. Since the present treatment is limited to 2-point functions,
it is sufficient to keep only bilinear terms in the gauge fields. 
In this approximation the action for the $SO(6)$ gauge fields reduces 
to a 
sum of uncoupled abelian sectors and it is convenient to use the 
St\"uckelberg formalism involving gauge invariant 
combinations $B_{\mu} = A_{\mu} + \pa_{\mu} \alpha$, 
with $\alpha$ bulk Goldstone fields. The corresponding 
field strengths are simply 
$F_{\m \n}= \pa_\m B_\n - \pa_\n B_\m$. (See \cite{Anatomy, bs2} 
where the
derivation of the vector sector Lagrangians from the full $D=5$ 
gauged supergravity theory is discussed in detail.) 
The bulk Lagrangian that describes this system of fields is given by
\be
S = S_0 +
\int_M \de^5x \sqrt{G} \left[{1 \over 4} K(\Phi) 
F_{\k \l} F_{\m \n} G^{\k \m} G^{\l \n} + \half M^2(\Phi) 
G^{\m \n} B_\m B_\n \right]
\label{action}
\ee
where $S_0$ is given in (\ref{scaction}) and 
$K(\Phi)$ and $M^2(\Phi)$ are positive semi-definite 
functions of the active scalar $\F$.  
We will only need their value along the solutions of the field 
equations.
For the GPPZ flow, up to a numerical rescaling, the kinetic term is 
canonical
and 
\be
M^{2} = 
{\sinh}^{2}\left({2 \F \over \sqrt{3}}\right) = {(1-u) \over u^2}.
\label{gppzvecmass}
\ee
For the CB flow
\bea \label{Kfun}
K_{SO(4)} &=& e^{{4\over \sqrt{6}}\F} = v^{2/3}
\\
K_{SO(2)} &=& e^{-{8\over \sqrt{6}}\F} = v^{-4/3}
\\
K_{X} &=& e^{-{2\over \sqrt{6}}\F} = v^{-1/3}
\eea
where $X$ denotes the coset $SO(6)/SO(4)\times SO(2)$.
The AdS masses vanish for the ``unbroken'' part of the R-symmetry 
group
$M_{SO(4)}^{2} = M_{SO(2)}^{2} = 0$, while 
\be
M_{X}^{2} = 4
{\sinh}^{2}\left({\sqrt{6}\F \over 2}\right) = {(1-v)^{2} \over v}.
\label{cbvecmass}
\ee

The field equations that follow from this action are
\bea 
&&\nabla_\m \left(K(\F) F^{\m \n}\right)
= M^2(\F) B^{\n} \ , \label{geqn} \\
&&\Box_G \Phi = {\pa V \over \pa \F}  
+ {1 \over 4} {\pa K \over \pa \F} 
G^{\k \m} G^{\l \n} F_{\k \l} F_{\m \n} + \half 
{\pa M^2 \over \pa \F} G^{\m \n} B_\m B_\n \ , \label{seqn} \\
&&R_{\m \n} = 
- 2 \left[T_{\m \n} - {1 \over 3} G_{\m \n} \Tr T \right], 
\label{greqn}
\eea
where $T_{\m \n}$ is the matter stress energy tensor and 
$\Tr T = G^{\m \n} T_{\m \n}$. 

As in past work on holographic renormalization it is most convenient 
to work in the coordinate system where the bulk metric takes 
the form
\be
\label{coord}
ds^2 = G_{\m \n} dx^\m dx^\n = {d\r^2 \over 4 \r^2} + 
{1 \over \r} g_{ij}(x,\r) dx^i dx^j 
\ee
where\footnote{ \label{oddFG}
In the Fefferman-Graham framework the most general expansion may 
contain half-integrals powers of $\r$, or integral powers of the coordinate 
$U, \r=U^2$ (the $\r$ variable was introduced in \cite{HS}). 
In the case of pure gravity and all the cases we consider here,
the coefficients with odd powers of $U$ can be shown to vanish
(except the $g_{(d)ij}$ coefficient in $d+1$ dimensions with $d$ 
odd). See however \cite{berg} for an example in $d+1=3$
that involves a near-boundary expansion  with half integral 
powers of $\r$.}
\be
\label{metrexp}
g_{ij}(x,\r) = g_{(0)ij} + g_{(2)ij} \r + \r^{2} [g_{(4)ij} 
+ h_{1(4)ij} \log \r + h_{2 (4)ij} (\log \r)^2] + ... 
\ee
Any asymptotically AdS metric can be brought to be of this form 
near the boundary \cite{FeffermanGraham}. The radial variables $\r$ 
and
$r$ are related by $\r = \exp(-2r)$. The boundary 
is located at $\r=0$ and the regularized action will be defined by
restricting to 
$\rho \geq \e$.

In the coordinate system (\ref{coord}), the scalar field equation 
reads 
\bea
\label{nbsceq}
&&4 \r^2 \Phi'' - 4 \r \Phi' ( 1 - \half \r (\log{g})') 
+ \r \Box_g \Phi - {\pa V \over \pa \Phi} =  \\
&&\quad {\r^{2}\over 4} {\pa K \over \pa \Phi} 
[g^{mn} g^{kl} F_{mk} F_{nl}  + 8 \r g^{kl} F_{k\r} F_{l\r}]
+{\r\over 2}{\pa M^2 \over \pa \Phi} 
[g^{ij} B_{i} B_{j} + 4 \r B_{\r}B_{\r}]
\nonumber
\eea 
where prime indicates derivative with respect to $\r$, 
$g=\det g_{ij}(x,\r)$,
and  $\Box_g$ is the scalar Laplacian in the metric $g_{ij}(x,\r)$.

Keeping terms which are at most quadratic in the gauge 
fields and dropping R-symmetry indices,
the vector field equations (\ref{geqn}) read
\bea
\label{vecrad}
&&\partial_i ({\sqrt{g}} K g^{ij} F_{j\r}) 
= M^2 {\sqrt{g}\over \r} B_\r
\\
&&\partial_i ({\sqrt{g}} K g^{ij} g^{kl} F_{jl}) 
+ 4\r \partial_\r ({\sqrt{g}} K g^{kl} F_{\r l})
= M^2 {\sqrt{g}\over \r} g^{ki} B_{i} 
\label{vectrans}
\eea
Taking the divergence of (\ref{vectrans}) we obtain
\be
\partial_i ({\sqrt{g}} M^2 g^{ij} B_j) 
+ 4 \r^{2} \partial_\r ({\sqrt{g}\over \r} M^2 B_{\r})=0
\label{veclong}
\ee

The Einstein equations in the coordinate system (\ref{coord})
are given by
\bea 
\label{eqnmatter}
&&\rho \,[2 g^{\prime\prime}_{ij} - 2 (g^\prime g^{-1} g^\prime)_{ij} 
+ 
\Tr\,
(g^{-1} g^\prime)\, g^\prime_{ij} \,] + R_{ij}[g] + 2
g^\prime_{ij} - \Tr \,(g^{-1} g^\prime)\, g_{ij} =\nonu  
&&\qquad 
- 2 \left(
 \pa_i \Phi \pa_j \Phi + {2\over 3\rho} [V(\F) - V(0)] g_{ij} + M^2 B_i B_j 
 + \r K [g^{kl} F_{ik} F_{jl} + 4 \r F_{i\r} F_{j\r}] \right.
 \nonu
&&\left. \qquad - {1\over 6} \r K g_{ij} [g^{mn} g^{kl} F_{mk} 
F_{nl}  
+ 8 \r g^{kl} F_{k\r} F_{l\r}] \right) 
\label{eqnij} 
\\
&&\nabla_i \Tr \,(g^{-1} g^\prime) - \nabla^j g_{ij}^\prime= 
- 2 \left( \pa_i \Phi \pa_\r \Phi + M^2 B_i B_\r 
+ \r K g^{kl} F_{ik} F_{\r l} \right)
\label{eqnir}  
\\
&&\Tr \,(g^{-1} g^{\prime\prime}) - \frac{1}{2} \Tr\, (g^{-1} 
g^\prime 
g^{-1} g^\prime) =
- 2 \left( \pa_\r \Phi \pa_\r \Phi + {1\over 6\rho^2} [V(\F)-V(0)] + M^2 B_\r 
B_\r 
\right. \nonu
&&\left. \qquad  + {2 \over 3} \r K g^{kl} F_{k\r} F_{l\r}
- {1 \over 24} K g^{mn} g^{kl} F_{mk} F_{nl} \right) 
\label{eqnrr}
\eea

We are now ready to study the near-boundary solution of the above 
equations.


\section{Near-boundary analysis: generalities}
\setcounter{equation}{0}

In this section we present a general discussion of the 
near-boundary analysis. The emphasis is on the general 
structure rather than the details. In subsequent sections 
we specialize the analysis to the two specific flows
we described in section 2. Although it was useful to distinguish
between background fields and fluctuations in the previous discussion,
the near boundary analysis applies to the full bulk fields of the
action (\ref{action}) and must be later linearized with respect to
the background. 
\subsection{Asymptotic solutions}

The first step in the analysis is to find asymptotic solutions
of the bulk field equations. To simplify the notation we 
suppress all spacetime and internal indices and collectively
denote bulk fields by $\cf(x,\r)$. 
Near the boundary  each field has an asymptotic expansion of 
the form
\be 
\label{ansatz}
\cf(x,\r)=\r^m \left(f_{(0)}(x) + f_{(2)}(x) \r + ... + 
\r^n (f_{(2 n)}(x) + \log \r\ \tilde{f}_{(2 n)}(x) + ...)+ ... \right)
\ee 
The field equations are second order differential equations 
in $\r$, so there are two independent solutions. Their 
asymptotic behaviors are $\rho^{m}$ and $\rho^{m+n}$, respectively.
For bosonic fields in domain wall solutions dual to flows driven
by protected operators deforming  
$\cn =4$ SYM theory, $n$ and $2m$ are non-negative integers. 
The boundary field $f_{(0)}$ that multiplies the leading behavior, 
$\r^m$,
is interpreted as the source for the dual operator.
In the near-boundary analysis one solves the field equations
iteratively by treating the $\r$-variable as a small parameter. 
This yields {\it algebraic} equations for $f_{(2k)}$, $k< n$,
that uniquely determine $f_{(2k)}$ in terms of $f_{(0)}(x)$
and derivatives up to order $2k$. These equations leave
$f_{(2 n)}(x)$ undetermined. This was to be expected: the 
coefficient $f_{(2 n)}(x)$ is the Dirichlet boundary condition
for a solution which is linearly independent from the one 
that starts as $\r^m$. As we will shortly see, $f_{(2 n)}$
is related to the VEV of the corresponding operator. The logarithmic
term in (\ref{ansatz}) is necessary in order to obtain a
solution. It is related to conformal anomalies of the dual theory, and
it is also fixed in terms of $f_{(0)}(x)$.

As we have remarked, any solution which is asymptotically AdS can be 
brought to the $\r$-coordinate system near the boundary. 
There are, however, bulk diffeomorphisms that preserve the 
form of the coordinate system (\ref{coord}), but  
induce a conformal transformation at the boundary \cite{ISTY}.
These transformations are related to the conformal anomaly \cite{HS} 
and thus 
with local RG equations on the field theory side, as discussed in
a preliminary way in \cite{johanna,howtogo}. 
As we will discuss, correlation functions are  
expressed in terms of certain coefficients in (\ref{ansatz}).
It follows that the RG equations are encoded in the 
transformations of the coefficients under the 
bulk diffeomorphisms given in \cite{ISTY}. 
Such transformations for various bulk fields have
been given in \cite{ISTY,KS,johanna,howtogo}.

\subsection{Counterterms}

The asymptotic solution obtained above can be inserted in the
regulated action. A finite number of terms which diverge as 
$\e \rightarrow 0$ can be isolated, so that the on-shell action
takes the form
\be \label{reg}
S_{{\rm reg}}[f_{(0)};\e] =  \int_{\r=\e} \de^{4}x \sqrt{g_{(0)}} 
[\e^{-\nu} a_{(0)}
+\e^{-(\nu+1)} a_{(2)}
+...-\log \e \ a_{(2 \nu)} + \co(\e^0)] \nonumber
\ee
where $\nu$ is a positive number that only depends on the scale
dimension of the dual operator and $a_{(2k)}$ are local functions
of the source(s) $f_{(0)}$. The counterterm action is defined as
\be
S_{{\rm ct}}[\cf(x,\e);\e] 
=-{\rm divergent\ terms\ of\ } S_{{\rm reg}}[f_{(0)};\e]
\ee
where divergent terms are expressed in terms of the fields 
$\cf(x,\e)$ `living' at the regulated surface $\r=\e$
and the induced metric there, $\c_{ij} = g_{ij}(x,\e)/\e$. This is 
required
for covariance and entails an ``inversion'' of the expansions 
(\ref{metrexp}),(\ref{ansatz}) up to the required order.

To obtain the renormalized action we first define a subtracted action 
at the cutoff 
\be \label{renact}
S_{{\rm sub}}[\cf(x,\e);\e] = 
S_{{\rm reg}}[f_{(0)};\e] + S_{{\rm ct}}[\cf(x,\e);\e].
\ee
The subtracted action has
a finite limit as $\e \to 0$, and the renormalized action is a 
functional
of the sources defined by this limit, i.e.
\be \label{sren}
S_{\rm ren}[f_{(0)}] = \lim_{\e \rightarrow 0} S_{{\rm sub}}[\cf;\e]
\ee
The distinction between $S_{\rm sub}$ and $S_{\rm ren}$ is needed 
because,
as described in Sec \ref{One-point functions},
the variations required to obtain correlation functions are performed
before the limit $\e \rightarrow 0$ is taken.

The procedure above amounts to a ``minimal'' scheme in which the 
divergences
of $S_{\rm reg}$ are subtracted. 
As in standard quantum field theory, one still has the freedom 
to add finite invariant counterterms. These correspond to a change of 
scheme.
For example, such finite counterterms may be needed 
in order to restore some symmetry (\eg supersymmetry \cite{howtogo}).

Given a bulk action there is a universal set of counterterms 
that makes the on-shell action finite for {\it any} solution 
of the bulk field equations with the same Dirichlet boundary data. 
The counterterms are different for different bulk actions, for 
example for different potentials $V(\F)$. 

The counterterm action can be decomposed into a set of 
counterterms associated to each field plus cross-terms.  
The former are universal in the sense that they are present 
everytime the bulk action contains a given bulk field, and
the latter vanish if all bulk fields 
but one are set to their background value.  
This structure follows from the fact that the sources are independent. 
If one wants to study the 
theory with only one source turned one, one can
either set all other sources to zero from the beginning
and then compute counterterms or
first compute counterterms with other sources non-vanishing
and then set the other sources to zero.
Quite obviously the two procedures should lead to the same result.
In actual computations however, as the ones presented later on,
this structure of the counterterm action comes about in a very 
non-trivial way. 
The dependence of the divergences in $S_{{\rm reg}}$ on a given 
source changes when different number bulk fields are involved in the 
computation.
However, the relation between the sources and 
the induced fields also changes when more bulk fields
are involved in the computation.  
The combined effect is that the counterterms take the form discussed above.

If the bulk field equations are coupled then one has to study the 
asymptotic solutions of the coupled system of the equations 
in order to obtain the correct set of counterterms. In some
cases, however, the near-boundary equations decouple 
up to the order that determine coefficients that contribute to divergences.
In these cases it is sufficient to study each field separately
in a fixed background. 
This may simplify computations
if one is only interested in computing a subset of all possible
correlation functions.
An example of this is the case scalars coupled
to gravity in the Coulomb branch flow discussed in section 5.1. 
If one only wants to 
compute correlation functions of the operator dual to 
the active scalar, the near-boundary analysis in section 
5.1.1 is sufficient. Only if the one wants to study correlation
that involve the stress energy tensor as well one needs
to analyze the full set  of gravity-scalar equations.
This should be contrasted with the 
case of the scalars coupled to gravity in the GPPZ flow,
discussed in section 5.2. In that case (which is the generic one)
the gravity-scalar equations do not decouple in near-boundary 
approximation, so even when one is interested in 
only computing correlation functions of the operators dual 
to the active scalar, one still has to analyze the full 
set of the gravity-scalar equations.

\subsection{One-point functions} 
\label{One-point functions}

Having obtained the renormalized on-shell action one can compute 
correlation functions by functionally differentiating $S_{\rm{ren}}$
with respect to the sources. 
After the renormalization procedure discussed above variation of 
(\ref{action}) reads
\bea 
\delta S_{\rm{ren}}[g_{(0)ij}, \f_{(0)}, A_{(0)i}, a_{(0)}] &=& 
\int d^4x \sqrt{g_{(0)}} [\half \<T_{ij}\> \delta g_{(0)}^{ij} 
+ \<O_\F\> \delta \f_{(0)} \nonu
&&+ \< J^{i} \> \delta A_{(0)i}
+ \<O_\a \> \delta a_{(0)} ] \label{varact}
\eea
where $g_{(0)}^{ij}, A_{(0)i},\f_{(0)}, a_{(0)}$ are sources for the 
dual operators and appear as the leading 
coefficients in the near boundary expansions of 
the bulk metric $G_{\mu \nu}$, gauge field $A_\m$,
active scalar $\F$ and St\"{u}ckelberg 
field $\a$, respectively. 

The expectation value of any scalar operator, such as the dual to 
$\Phi$ 
is defined by
\be \label{1ptdef}
\< O_{\F} \> = {1 \over \sqrt{g_{(0)}}} 
{\delta S_{\rm{ren}} \over \delta \f_{(0)}}
\ee
It can be computed by rewriting it in terms of the fields living at 
the 
regulated boundary\footnote{For scalars of 
dimension 2, an additional $\log\e$ is needed in this formula,
see (\ref{onecbop}).}
\be \label{oexp}
\< O_{\Phi} \> = 
\lim_{\e \to 0} \left( {1 \over \e^{\D/2}} {1 \over \sqrt{\g}} 
{\delta S_{\rm{sub}} \over \delta \F(x,\e)} \right)
\ee
where $\gamma_{ij}(x) = g_{ij}(x,\e)/\e$ is the induced metric on the 
boundary 
and $\g = \det(\g_{ij})$. This expression allows us to compute the 
1-point function in terms of (undetermined) coefficients in the
asymptotic expansion of the bulk field $\F$. 

The expectation value of the stress-energy tensor of the 
dual theory is given by
\be \label{tij1}
\<T_{ij}\> = {2 \over \sqrt{ g_{(0)}}} 
{\d S_{\rm{ren}} \over \d g_{(0)}^{ij}}
=\lim_{\e \to 0} 
{2 \over \sqrt{g(x, \e)}} {\d S_{\rm{sub}} \over \d g^{ij}(x,\e)}  
=\lim_{\e \to 0}\left( {1 \over \e}\, T_{ij}[\c]\right)
\ee
where $T_{ij}[\c]$ is the stress-energy tensor of the theory 
at $\r=\e$. From the gravitational point of view this is the 
Brown-York stress energy tensor \cite{BY} supplemented by 
appropriated counterterms contributions \cite{BK}
\be \label{tij2}
T_{ij}[\c]=T^{\rm{reg}}_{ij}+T^{\rm{ct}}_{ij},
\ee
$T^{\rm{reg}}_{ij}$ comes from the regulated bulk action 
and it is equal to (see (3.6) in \cite{KSS})
\bea \label{regtij}
T_{ij}^{\rm{reg}}[\c]&=&-{1 \over 2} ({\cal K}_{ij} - {\cal K} 
\c_{ij}) \\
&=&-{1 \over 2}\,(-\pa_\e g_{ij}(x,\e) + g_{ij}(x,\e)\, 
\Tr [g^{-1}(x,\e) \pa_\e g(x,\e)] + {3 \over \e} g_{ij}(x,\e)) 
\nonumber
\eea
where ${\cal K}_{ij}$ is the extrinsic curvature tensor.
$T^{{\rm{ct}}}_{ij}$ is the contribution due to the counterterms. 
 
Similarly, the 1-point function of R-symmetry currents reads
\be \label{jone}
\< J^{i} \> = {1 \over \sqrt{g_{(0)}}}  {\d S_{\rm{ren}} \over \d 
A_{(0)i}}
=\lim_{\e \to 0} \left({1 \over \e^2} 
{1 \over \sqrt{\c}} {\d S_{\rm{sub}} \over \d A_{i}(x,\e)}\right)
\ee

The 1-point functions $\< {\cal O}_{\F} \>, \< T_{ij} \>$ and $\< 
J_{i} \>$
depend on all sources. Field theory vacuum expectation values can be 
obtained 
by setting the sources to zero. The same remarks hold for 
$\< {\cal O}_{\a} \>$. However, it is somewhat formal to speak of a 
field
theory operator ${\cal O}_{\a}$ dual to a Stuckelberg scalar field. 
It is
better to convert to a linear formulation of the corresponding 
symmetry, and
this is done in the examples discussed in Sec. \ref{vecsec} 
   
\subsection{Two-point functions} 
\label{2pt}

In order to obtain higher point functions 
we need to functionally differentiate 1-point functions
with respect to the sources. 

For example one can define and compute the connected 2-point functions
of the stress tensor as
\be
\< T_{ij}(x)T_{kl}(y) \> = - {2 \over \sqrt{g_{(0)}(y)}}
{\d \< T_{ij}(x) \> \over
\d g_{(0)}^{kl}(y)} ,
\ee
and the connected 2-point function of the currents as
\be
\< J_{i}(x)J_{k}(y) \> = - {1 \over \sqrt{g_{(0)}(y)}}
{\d \< J_{i}(x) \> \over \d A_{(0)}^{k}(y)} .
\ee
In terms of the projectors  
\be
\pi_{ij} =\d_{ij} - {p_ip_j\over p^2}, \qquad
\P^{TT}_{ijkl} =
{1 \over 2} (\p_{ik} \p_{jl} + \p_{il} \p_{jk})-{1 \over 3} \p_{ij} 
\p_{kl}
\; 
\ee
the (Fourier transform of the) former decomposes into
\be
\<T_{ij}(p)T_{kl}(-p)\> =\P^{TT}_{ijkl} A(p^2) +\pi_{ij} \pi_{kl} 
B(p^2)
\ee
while the latter into
\be
\<J_{i}(p)J_{k}(-p)\> =\pi_{ik} C(p^2) + {p_{i} p_{k}\over p^{2}} 
D(p^{2})
\ee
Supersymmetry Ward identities are expected to 
relate $A$ to $C$ and $B$ to $D$, but we shall not derive 
these relations here.

\subsection{Ward identities}

Bulk gauge fields couple to boundary symmetry currents. It follows
that bulk gauge invariance translates into Ward identities of the 
boundary quantum field theory. We derive here the Ward identities
that the 1-point functions should satisfy. As we will explicitly
show in the next few sections, the holographically computed 
1-point functions do satisfy these Ward identities including 
anomalies.

Using (\ref{varact}), invariance of (\ref{action}) under 
diffeomorphisms,
\bea
&&\delta g_{(0)}^{ij} = - (\nabla^i \xi^j + \nabla^j \xi^i), \quad
\delta \f_{(0)} = \x^i \nabla_i \f_{(0)}, \nonu
&&\delta a_{(0)} = \x^i \nabla_i a_{(0)}, \quad
\delta A_{(0)i} = \x^j \nabla_j A_{(0) i} + \nabla_i \xi^j A_{(0)j},
\eea
implies the Ward identity for the conservation of the stress 
tensor\footnote{
Notice that these Ward identities are valid in the presence of 
sources. In particular, $\<T_{ij}\>$ depends on sources.
In field theory one usually expresses the Ward identities 
in terms of the stress energy tensor with the sources set 
equal to zero,  $\<T_{ij}\>_{QFT}=\<T_{ij}\>|_{{\rm sources}=0}$.},
\be \label{divW}
\nabla^i \<T_{ij}\> = -  \<O_{\F} \> \nabla_j \f_{(0)} 
-\<O_{\a} \> \nabla_j a_{(0)} 
- F_{(0)ij} \<J^{i}\> + A_{(0)j} \nabla_{i} \< J^{i} \> 
\ee
where $F_{(0)ij}$ is the field strength of $A_{(0)i}$.

Invariance under Weyl transformations,
\bea
&&\delta g^{ij} = - 2 \s g^{ij}, 
\qquad \delta \f_{(0)}=-(4-\D) \s \f_{(0)} \nonu
&&\d A_{(0)i} = - \s A_{(0)i}, \qquad \d a_{(0)} = - \s a_{(0)}
\eea
leads to the conformal Ward identity 
\be \label{trW}
\<T^i_i\>= - (4-\Delta) \f_{(0)} \< O_{\F} \> -a_{(0)} \< O_{\a} \> 
- \<J^i\> A_{(0)i}+{\cal A}.
\ee
where we have allowed for the conformal anomaly ${\cal A}$.
As explained in \cite{HS} ${\cal A}$ is obtained directly from the 
logarithmic counterterm of the the bulk action. 
 
In the case of the GPPZ flow, the bulk St\"{u}ckelberg
gauge invariance implies a corresponding Ward identity.
We will find convenient in section 6.2 to use normalizations
such that $B_{\mu}=A_\mu - 3\delta_{\mu} \a/2$. Then the bulk gauge 
invariance implies for the sources 
\be \label{Stuck}
\delta A_{(0)}^{i} = \nabla^i \lambda, \qquad
\delta a_{(0)} = {2 \over 3} \lambda
\ee
It follows 
\be
\nabla^{i}\< J_{i} \> = {2 \over 3} \<O_\a\>.
\label{currward}
\ee
A more detailed derivation of this Ward identity is given in section 
6.2.

In the Coulomb branch case, the renormalized on-shell is not invariant under 
transformations analogous to (\ref{Stuck}) because of the 
presence of Goldstone poles. We refer to section 6.1 for a 
detailed discussion of this point.


\section{Gravitational sector}

\subsection{Near-boundary analysis for the CB flow} 

The coordinate transformation that brings the domain-wall 
metric (\ref{dmw}) to the coordinate system (\ref{coord})
admits an expansion as \footnote{The exact coordinate transform is
$$ \ell^2 \rho=-3 \sqrt{3} \exp \left[{\p \over \sqrt{3}} 
- \sqrt{3} \tan^{-1} {1 \over \sqrt{3}} (1+2 v^{1/3})\right]
{(v^{1/3}-1) \over \sqrt{1 + v^{1/3} + v^{2/3}}}$$.}
\be \label{vrho}
1-v=\ell^2 \r -{2 \over 3} \ell^4 \r^2 + \co(\r^3)
\ee
In these coordinates the solution for $\Phi$ and $A$,  
is given by
\be \label{CB}
\Phi
={1 \over \sqrt{6}}(-\ell^2 \rho + {1 \over 6} \ell^4 \r^2 + 
\co(\r^3)), \ \
e^{2 A} = {1 \over \r}(1 -{1 \over 18}\ell^4 \r^2 + \co(\r^3)) 
\nonumber 
\ee
The active scalar has dimension $\Delta=d/2=2$ and can be decomposed as
\be 
\label{cbscalexp}
\F(x,\r)=\r \log\r \, \f(x,\r) + \r \tf(x,\r)  \label{F} 
\nonumber
\ee

\subsubsection{Scalars in a fixed background} 
\label{nogr}

Let us first study the case of a scalar
field in a given background of the form (\ref{coord}). 
As we will see in the next section, the backreaction of the 
scalar field to the gravitational field equations does not 
affect the coefficient of the metric that determine the 
divergences of the on-shell action, and vice versa.
This means that for the purpose of the near-boundary 
analysis the scalar equation decouples from Einstein's
equations, so we can consistently study scalars on a 
fixed background. This means that the results of this section 
are also part of the analysis of the next section.

The relevant field equation 
to solve is (\ref{seqn}) with the potential in (\ref{CBpot}).
We look for a solution of the form (\ref{cbscalexp})
where
\bea \label{phiexp}
&&\f(x,\r)=\f_{(0)} + \f_{(2)} \r + \r \log\r \psi_{(2)} + ..., 
\\
&&\tf(x,\r)=\tf_{(0)} + \tf_{(2)} \r + ... \nonumber
\eea
$\f_{(0)}$ is the source for a composite operator of dimension 
$\D=2$, and $\tf_{(0)}$ is proportional to the VEV of this operator 
(as we will shortly derive). 

The field equation (\ref{seqn}) can be solved order by order in the 
$\rho$ variable.
One obtains
\bea \label{feqs}
&&\f_{(2)}=-{1 \over 4} \left( \Box_{(0)} \f_{(0)} 
+ {2 \over 3} \f_{(0)} R[g_{(0)}] \right) 
- {4 \over \sqrt{6}} (\f_{(0)}^2 - \half \f_{(0)} \tf_{(0)}) \nonu
&&\tf_{(2)}=-{1 \over 4} \left( \Box_{(0)} \tf_{(0)} 
+ {1 \over 3} R[g_{(0)}] (\tf_{(0)} + \f_{(0)}) + 8 (\f_{(2)} + 
\psi_{(2)})  
\right)+ {1 \over \sqrt{6}} \tf_{(0)}^2 \nonu
&&\psi_{(2)}= {1 \over \sqrt{6}} \f_{(0)}^2
\eea
As anticipated, $\tf_{(0)}$ is not determined
by these equations.

Following the procedure outlined above, the regularized action is 
given by 
\bea \label{matreg}
S_{\rm{reg}}&=& \int_{\r \geq \e} \de^5 x\, \sqrt{G}
\left( \half G^{\m \n} \pa_\m \F \pa_\n \F
-2 \F^2  +{4 \over 3 \sqrt{6}} \F^3 \right) + ... \nonu
&=&-\int \de^d x \sqrt{g_{(0)}} \left[ \log^2 \e \f_{(0)}^2
+ \log\e (\f_{(0)}^2 + 2 \f_{(0)} \tf_{(0)}) 
+ {\cal O} (\epsilon^{0}) \right] \nonu
&=&-\int \de^4 x \sqrt{\g}
\left[\Phi^2(x,\e) + {\Phi^2(x,\e) \over  \log \e}\right]  
+ {\cal O} (\epsilon^{0})
\eea
Notice that the counterterm proportional to $1/\log\epsilon$ 
is not vanishing in the UV, i.e. in the $\e \to 0$ limit. 
This is so because it is $\f_{(0)}(x)$  rather than $\Phi(x, \e)$ 
that 
is kept fixed in this limit. By inserting the expansion of 
$\gamma(x,\e)$ 
and $\F(x,\e)$ into the last expression in (\ref{matreg}) one easily 
verifies
that the infinite terms reproduce that of the second expression in 
(\ref{matreg}). The latter, however, differs from the 
former in its finite part. 

The renormalized action is given by 
\be \label{renorm}
S_{\rm{ren}}=\lim_{\e \to 0} S_{\rm sub} \equiv \lim_{\e \to 0}
\left[ S_{\rm{reg}} 
+ \int \de^4 x \sqrt{\g}
\left({\D \over 2} \Phi^2(x,\e) + {\Phi^2(x,\e) \over \log\e}
\right) \right] 
\ee
It is important that the counterterms are expressed
in terms of field living on the regulating hypersurface.
Otherwise, the subtraction would not be covariant. 
Although we have presented the derivation for $D=4+1$ bulk dimensions,
the final result is valid for operators of dimension $\D=d/2$ in 
(bulk)
dimension $D=d+1$. The same holds for most of the following results 
in this 
subsection.

Let us now compute the 1-point function,
\be 
\<O_{\F} \> = {1 \over \sqrt{g_{(0)}}} {\d S_{\rm{ren}} \over 
\d\f_{(0)}}=
\lim_{\e \rightarrow 0}\left( 
{\log \e \over \e} {1\over\sqrt{\gamma}}
{\d S_{\rm{sub}} \over \d\F(x,\e)} \right)
\label{onecbop}
\ee
where the first equality is the definition of the 1-point function 
and in the second equality we express things in terms of the regulated
theory at $\r=\e$. Using (\ref{renorm}) we obtain
\be
{\d S_{\rm{sub}} \over \d\F(x,\e)}
= 2 \left(- \e \partial_\e \F(x,\e) + {\D \over 2} 
\F(x,\e) + {\F(x,\e) \over \log\e}
\right)
\ee
where the last two terms is the contribution due to counterterms.
Inserting in (\ref{onecbop}) one obtains
\be \label{oexpe}
\< O_\F \> = 2 \tf_{(0)}
\ee
This is the relation advocated in \cite{KleWit}\footnote{For $\D  =2$ 
scalar
operators, the factor $2\D -4$ is replaced by 2.}. From this relation 
we conclude that the solution in (\ref{CB}) indeed describes a 
different vacuum state rather than a deformation of the original theory.
Setting $\tf_{(0)} = -1/\sqrt{6}$, its background value, one finds the
physical $ \<\co_\F\> = -2/\sqrt{6}$ of this Coulomb deformation of 
the
$\cn =4$ theory. We view the derivation of vevs as a clean and elegant
application of holographic renormalization.
Had one differentiated the regulated action (5.33) wrt
$\phi_{(0)}$ and then discarded infinities, one would have gotten an 
incorrect 
result. This is so because the covariant counterterms include finite 
parts, 
as already remarked.

\subsubsection{Scalars coupled to gravity} 
\label{wigr}

We now turn to study the backreaction of the scalars on the 
gravitational sector. To solve the equations we insert the 
asymptotic expansions in Einstein's equations and solve them
order by order in the $\r$-variable. The most efficient way of 
performing this computation is to first differentiate the 
equations with respect to $\r$ and then set $\r$ equal to zero.
For instance, if we want the equations that appear at order 
$\r^2$, we may differentiate the equations twice with respect to $\r$ 
and then
set $\r=0$. The resulting equations are algebraic and can be readily
solved to obtain the coefficients in the asymptotic expansion of the 
bulk
fields.

To leading order, one sets $\r=0$ in Einstein's equations, 
solves them and finds that the scalars do not contribute and 
$g_{(2)}$ obtains the same value as for the case of pure gravity,
\be \label{g2}
g_{(2)}{}_{ij} =  \frac{1}{2} \left( R_{ij} - \frac{1}{6} 
R\, g_{(0)ij} \right).
\ee
Here and henceforth curvature tensors and index contractions pertain to 
$g_{(0)ij}$.

The next order equations determine 
$h_{1 (4)}, h_{2 (4)}$, the trace of $g_{(4)}$ and the covariant
divergence of $g_{(4)}$,
\bea
&&h_{1(4)ij}=h_{(4)ij} - {2 \over 3} \f_{(0)} \tf_{(0)} g_{(0)ij}, 
\qquad h_{2(4)ij}=-{1 \over 3} \f_{(0)}^2 g_{(0)ij} \label{h} \\
&& \Tr g_{(4)} = {1 \over 4} \Tr g_{(2)}^2 
-{2 \over 3} (\f_{(0)}^2 + 2 \tf_{(0)}^2), \label{trace}  \\
&& \nabla^j g_{(4) ij} = \nabla^j 
\left(- {1 \over 8}[\Tr\, g_{(2)}^2 - (\Tr\, g_{(2)})^2]\, g_{(0) ij} 
+ \half (g_{(2)}^2)_{ij} - {1 \over 4}\, g_{(2) ij}\, \Tr\, g_{(2)} 
\right. \nonu 
&&\left. \qquad \qquad
-{1 \over 3} (2 \f_{(0)}^2 + \tf_{(0)}^2) g_{(0)ij}\right) 
- (\tf_{(0)} \nabla_i \f_{(0)} -\f_{(0)} \nabla_i \tf_{(0)})
\label{div}
\eea
where
\bea 
h_{(4)ij}&=&{1 \over 8} R_{ikjl} R^{kl} + {1 \over 48} \nabla_i 
\nabla_j R
-{1 \over 16} \nabla^2 R_{ij} -{1 \over 24} R R_{ij} \nonu
&&+ ({1 \over 96} \nabla^2 R + {1 \over 96} R^2 -{1 \over 32} 
R_{kl}R^{kl}) 
= \half T_{ij}^a \label{h4}
\eea
and $T_{ij}^a$ is the stress energy tensor derived from the action 
\be
S_a=\int d^4x \sqrt{g_{(0)}} \ca_{grav} \ .
\ee
$\ca_{grav}$ is the gravitational trace anomaly given in 
(\ref{gravanom}).
The solution of the scalar field equation is described in the 
previous subsection and the results given 
(\ref{cbscalexp})-(\ref{phiexp})-(\ref{feqs})
carry over.

Having obtained the asymptotic solution we proceed to calculate the 
regularized action. 
\be
S_{\rm{reg}}= \int_{\r\geq \e} \de^5x \sqrt{G} \left[{1 \over 4} R  
+\half G^{\m \n} \pa_\m \F \pa_\n \F + V(\F) \right]
-\half \int_{\r=\e} \sqrt{\c} {\cal K} \label{regact} \\
\label{cbregact}
\ee
where $\gamma$ is the induced metric at $\r = \e$.
A direct computation yields
\bea 
\label{infin}
S_{\rm{reg}}&=&\int_{\r \geq \e} \de \r \de^4 x 
{\sqrt{g(x,\r)} \over 2 \r^{3}} \left[- {2 \over 3} V \right] - 
{1 \over 2} \int_{\r = \e} \de^4 x \sqrt{\gamma} \ck \\
&=& \int \de^4 x 
\sqrt{g_{(0)}} \left[
-{3 \over 2 \e^2} - \log\e 
{1 \over 8} [(\Tr g_{(2)})^2 - \Tr g_{(2)}^2] 
- \log\e \phi_{(0)}^2 \right] \nonumber
\eea
In the first equality we used Einstein's equations to eliminate the 
scalar 
curvature from the action, and in the second equality we have 
inserted 
the asymptotic 
solution we just obtained. The logarithmic divergence agrees exactly 
with
the one determined in (\ref{matreg}), and the gravitational one
with the results of \cite{HS} (as they should).

In order to compute the renormalized action (\ref{renact}), the 
divergent 
terms (as $\e \to 0$) in (\ref{regact}) are to be cancelled by 
adding counterterms $S_{\rm{ct}}=-S_{\rm{div}}$.
When expressed in terms of the fields
living at $\r=\e$, the result is 
\be \label{ctgf}
S_{\rm{ct}}=\int_{\r=\e} \de^4 x \sqrt{\g} \left({3 \over 2} 
-{1 \over 8} R 
-\log\e {1 \over 32} (R_{ij} R^{ij} - {1 \over 3} R^2)
+ \Phi^2(x,\e)  + {\F^2(x,\e) \over \log\e} \right).  
\ee
To derive this result one needs to 
invert the relations between the induced fields $\g_{ij}(x,\e)$, 
$\F(x,\e)$ and the sources $g_{(0) ij}$ and $\f_{(0)}$. Details
on how to do this are given in appendix B of \cite{KSS}.
The simplest way to check (\ref{ctgf}) is to to insert the expansions 
of $\gamma$ and $\Phi$ and show that the divergent terms agree
with the ones in (\ref{infin}). Notice that the $\log^2 \e$ 
divergence 
due to  the $\Phi^2$ term cancels against a similar divergence 
originating from $\sqrt{\g}$.
The first three counterterms in (\ref{ctgf}) are the gravitational 
counterterms 
(see (B.4) of \cite{KSS}). 
Notice that the matter dependent infinities in (\ref{infin}) 
are different from the ones in (\ref{matreg}). Nevertheless, 
the required matter counterterms  
coincide with the ones we computed in the previous section.
The difference between the infinite parts is exactly compensated 
by the backreaction of the scalars to the metric.

We can now proceed to compute the renormalized stress tensor.
The contribution due to counterterms is given by
\be \label{counterT}
T^{\rm{ct}}_{ij}=-{3 \over 2} \c_{ij} 
- {1 \over 4} (R_{ij}[\g] 
- \half R[\g] \c_{ij}) - \g_{ij} \left(\Phi^2(x,\e) 
+ {\Phi^2(x,\e) \over \log\e}\right) -
\half \log\e  T_{ij}^a
\ee
Combining (\ref{regtij}) and (\ref{counterT}) we obtain
\bea
&&\<T_{ij}\>=-{1 \over 2} \lim_{\e \to 0}
\left[{1 \over \e} (-g_{(2)ij} + g_{(0)ij} \Tr\, g_{(2)}
+ \half R_{ij} - {1 \over 4} g_{(0)ij} R)
\right. \nonu
&&\hspace{-1cm}+\log^2 \e \left( -2 h_{2(4) ij} - {2 \over 3} 
\f_{(0)}^2 g_{(0)ij}\right) \nonu
&&\hspace{-1cm}+\log\e\, \left(-2 (h_{1(4)ij} +h_{2(4)ij}) + T^a_{ij} 
-{2 \over 3} (2 \f_{(0)} \tf_{(0)} +  \f_{(0)}^2) g_{(0)ij} \right) 
\nonu
&&\hspace{-1cm}
-2 g_{(4)ij} - h_{1(4)ij} + g_{(2)ij} \Tr\, g_{(2)} - \half g_{(0)ij} 
\Tr\, g_{(2)}^2  \\
&&\hspace{-1cm}+{1 \over 8}( R_{ik} R^k{}_{j}
-2 R_{ikjl} R^{kl} -{1 \over 3} \nabla_i \nabla_j R + \nabla^2 R_{ij}
-{1 \over 6} \nabla^2 R g_{(0) ij})  \nonu
&&\left. \hspace{-1cm}-{1 \over 4} g_{(2)ij} R + {1 \over 8} g_{(0)ij}
(R_{kl} R^{kl} -{1 \over 6} R^2) 
-{2 \over 3}(2 \f_{(0)}^2 +  \tf_{(0)}^2 
- 2 \f_{(0)} \tf_{(0)}) g_{(0)ij}\right]. \nonumber
\eea
where the curvature tensor is that of $g_{(0)}$.
Using (\ref{g2})-(\ref{h}) we find that the $1/\e$, 
$\log\e$ and $\log^2 \e$ divergences cancel. The finite part 
is the expectation value of the stress tensor in the dual QFT. It is equal to 
\bea \label{cbT}
\<T_{ij}\>&=& g_{(4)ij} + 
{1 \over 8}[\Tr\, g_{(2)}^2 - (\Tr\, g_{(2)})^2]\, g_{(0) ij} 
- \half (g_{(2)}^2)_{ij} \\
&+& {1 \over 4}\, g_{(2) ij}\, \Tr\, g_{(2)} 
+{1 \over 3} (\tf_{(0)}^2 - 3\f_{(0)} \tf_{(0)}) g_{(0)ij} 
+{2 \over 3} \f_{(0)}^2 g_{(0)ij} 
+ {3 \over 2} h_{(4) ij} \nonumber
\eea
Notice that because $h_{(4)ij}$ is equal to $\half T^a_{ij}$ 
the contribution in the boundary stress energy tensor proportional 
to $h_{(4)ij}$ is scheme dependent. Adding a local finite counterterm 
proportional
to the gravitational trace anomaly will change the coefficient of 
this term. 
Similar remarks apply to the penultimate term:
it is proportional to the stress energy 
tensor derived from an action equal to the matter
conformal anomaly (\ref{cbscalanom}). 
One may remove such contributions from the boundary stress
energy tensor by a choice of scheme.

Even though $g_{(4)ij}$ is not fully determined by the
bulk field equations, its divergence and trace are, and this is 
sufficient
in order to compute the trace and divergence
of $\<T_{ij}\>$ given in (\ref{cbT}). A direct computation using the 
asymptotic solution yields,
\bea
\nabla^i \<T_{ij}\> &=& -  \< O_\F\> \nabla_j \f_{(0)} \label{divWCB} 
\\
\<T^i_i\>&=&- 2 \f_{(0)} \< O_\F \>
+{1 \over 16} (R_{ij} R^{ij} -{1 \over 3} R^2)
+ 2\phi_{(0)}^2  \qquad \label{trWCB}
\eea
i.e. $ \<T_{ij}\>$ correctly
satisfies the diffeomorphism and trace Ward identities.
The last two terms in (\ref{trWCB}) are what we called $\ca$ in
(\ref{trace}). The second term
\be
{\ca}_{grav} = {1 \over 16} (R_{ij} R^{ij} -{1 \over 3} R^2)
\label{gravanom}
\ee
is the holographic gravitational
conformal anomaly \cite{HS} and the last term
\be
{\ca}_{scal} = 2\phi_{(0)}^2
\label{cbscalanom}
\ee
is the conformal
anomaly due to matter \cite{PeSk}. The coefficients in both
of them are known not to renormalize, and indeed we obtain
the correct value. The Ward identities and the anomalies
are important checks of the intermediate computations and of the
consistency of the formalism.


\subsection{Near-boundary analysis for the GPPZ flow}

In the coordinate system (\ref{coord}), in which   
$\rho=1-u$, the functions
$\Phi$ and $A$ that determine the GPPZ domain wall solution expand as
\be \label{gppzwall}
\Phi = \rho^{1/2} (\sqrt{3}  + {1\over \sqrt{3}} \rho + \co
(\rho^{2})), \quad e^{2A} = {1\over \rho} - 1
\ee
Similarly, the expansion of the superpotential reads
\be
W = - {3\over 2} - {3 \over 2} \rho + \co (\rho^{2})
\ee
In our units, the quadratic term in the potential has $m^{2} = - 3$ 
implying that the operator dual 
to $\Phi$ has scaling dimension $\Delta = 3$.

The gauge kinetic function of the $U(1)$ graviphoton is $K=3$
and its effective mass term \cite{Anatomy} is 
\be
M^{2} = - 2 A'' = {4} {\rho \over (1-\rho)^{2}} \approx 
{4} [\rho + 2 \rho^{2} + \co (\rho^{3})].
\ee

\subsubsection{Scalars coupled to gravity} 
\label{gppzscalar}

In this section we perform the asymptotic analysis of the 
scalar-graviton
system. These results have been used in \cite{howtogo} without a 
proof.

The asymptotic expansion of the metric is given in (\ref{coord}). The 
expansion of the scalar field reads
\be
\Phi(x,\rho) = \rho^{1/2} \left( \phi_{(0)}(x) + \rho \phi_{(2)}(x) + 
\rho \log{\rho} \psi_{(2)}(x) + \ldots \right)
\label{gppzscalexp}
\ee

Plugging (\ref{gppzscalexp}) into the scalar field equation 
one gets 
\be
\psi_{(2)} = -{1\over 4} \nabla^{2} \phi_{(0)} + {1\over 6} 
R[g_{(0)}] \phi_{(0)}
\ee
while the dependence of $\phi_{(2)}$ on the sources $\phi_{(0)}$
and $g_{(0)}$ remains undetermined.

Einstein's equations to the lowest order yield
\be \label{g2gppz}
{g}_{(2)}{}_{ij} =  \frac{1}{2} \left( R_{(0)ij} - \frac{1}{6} 
R_{(0)} \, g_{(0)ij} \right) - {1\over 3} \phi_{(0)}^{2} g_{(0)ij}
\; .
\ee
{}From the computation of divergences in pure gravity we know 
that gravity divergences depend on $g_{(2)}$. Since there 
is a backreaction of the scalar to $g_{(2)}$, these divergent
terms lead also to scalar dependent divergences. These terms
would be missed in a fixed background computation where 
$g_{(2)}$ is fixed and independent of the scalar fields.
This is an example of the general discussion at the end 
of section 4.2. In cases like this, one needs to solve
the coupled system of equations. Notice that 
in the Coulomb branch case the scalar field
did not backreact on any of the metric coefficient that 
contribute to divergences, so it was consistent to 
consider the problem on a fixed gravitational background.

The non logarithmic terms in Einstein equations yield
\bea 
&&\Tr g_{(4)} = -2 \f_{(0)} \f_{(2)} - {1 \over 4}  \f_{(0)} \Box_0 
\f_{(0)}
-{5 \over 72} R \f_{(0)}^2 \nonu
&&\qquad \qquad
+ {1 \over 16}(R_{ij} R^{ij} -{2 \over 9} R^2) + {2 \over 9} 
\f_{(0)}^4 \\
&&\nabla^j g_{(4)ij}= \nabla^j
\left(- {1 \over 8}[\Tr\, g_{(2)}^2 - (\Tr\, g_{(2)})^2]\, g_{(0) ij}
+ \half (g_{(2)}^2)_{ij}
\right. \nonu
&&\qquad \qquad
- {1 \over 4}\, g_{(2) ij}\, \Tr\, g_{(2)}
-{3 \over 2} h_{1(4)ij} - \half \nabla_i \f_{(0)} \nabla_j \f_{(0)}
 \nonu
&&\left.\qquad \qquad
+ g_{(0) ij}\left[{1 \over 4} (\nabla \f_{(0)})^2 - \f_{(0)} (\f_{(2)}
+ \psi_{(2)})\right]  \right) \nonu
&&\qquad \qquad
- [-2 (\f_{(2)} + \psi_{(2)}) + {2 \over 9} \f_{(0)}^3] \nabla_i 
\f_{(0)}
\eea
The other coefficient tensor ${h}_{1(4)ij}$ in the expansion of the
GPPZ metric follows from the logarithmic terms in the coupled field 
equations.
Combining with the purely gravitational result ${h}_{(4)ij}$,
defined in (\ref{h4}),
that is both transverse and traceless, one gets
\bea
&&h_{1(4)ij}=h_{(4)ij}+{1 \over 12} R_{ij} \f_{(0)}^2
-{1 \over 3} \nabla_i \f_{(0)}\nabla_j \f_{(0)}
+{1 \over 12} (\nabla \f_{(0)})^{2} g_{(0)ij} \nonu
&&\qquad \qquad
+{1 \over 6} \f_{(0)} \nabla_i \nabla_j \f_{(0)}
+ {1 \over 12} \f_{(0)} \Box_0 \f_{(0)} g_{(0)ij} \nonu
&&\qquad \qquad = h_{(4)ij} + {1\over 2} T^{\f}_{ij}
+ {1 \over 4}
g_{(0)ij} (\f_{(0)} \Box_0 \f_{(0)} + {1 \over 6} R \phi_{(0)}^2) 
\label{hgppz}
\eea
where $h_{ij}$ is given by (\ref{h4}) and 
$T^{\f}_{ij}$ is the stress energy tensor derived from the
action
\be
S_\f = \int \de^4x \sqrt{g_{(0)}} \ca_{scal}
\ ,
\ee
where $\ca_{scal}$ is the matter conformal anomaly 
given in (\ref{matan}).
Furthermore ${h}_{2(4)ij} =0$ in the GPPZ case.

Using the asymptotic solution one can calculate the 
on-shell regularized action as in (\ref{cbregact})
\bea
&&S_{\rm{reg}} = \int \de^4 x \sqrt{g_{(0)}} 
\left[ -{3\over 2\e^2} + {1 \over 2 \e} \f_{(0)}^2 
+ \log\e \left( {1 \over 32}
(R_{ij}[g_{(0)}] R^{ij}[g_{(0)}] - {1 \over 3} R^2[g_{(0)}])
\nonumber \right. \right.\\
&&\left. \left.
+ {1 \over 8} (\f_{(0)} \Box_{0} \f_{(0)} + {1 \over 6} R[g_{(0)}] 
\f_{(0)}^2)
\right) + \co(\e^0) \right]
 \label{gppzreg}
\eea
To obtain the counterterm action we need to rewrite the divergences 
in terms 
of the induced fields $\g_{ij}$ and $\F(x,\e)$. This amounts to 
formally 
inverting the asymptotic series expressing  $\g_{ij}$ and $\F(x,\e)$
in terms of $g_{(0)}$ and $\f_{(0)}$ to obtain the sources in terms 
of the induced fields. Inserting this result in (\ref{gppzreg})
(and including an overall minus) yields
\bea \label{gppzct}
&&S_{\rm{ct}} = \int_{\r=\e} \de^4 x \sqrt{\g}
\left( {3\over 2}  - {1 \over 8} R[\g]  + \half \Phi^2(x,\e) + 
{1\over 18} 
\Phi^4(x,\e) \right.
\\
&& -  \left. \log\e \left[ {1 \over 32} (R_{ij}[\g] R^{ij}[\g] - {1 
\over 3} 
R[\g]^2) +  {1\over 4}  
[\F(x,\e) \Box_{\gamma} \F(x,\e) + {1\over 6} R \F^2(x,\e)]\right] 
\right).  
\nonumber
\eea
The quartic term in $\Phi$, which is actually finite, is 
introduced to preserve supersymmetry \cite{howtogo}. 

We can now proceed to compute the 1-point functions.
For the scalar operator we obtain
\be
\< O_\F \> = {1 \over \sqrt{g_{(0)}}} 
{\delta S_{\rm{ren}} \over \delta \f_{(0)}}
= \lim_{\e \to 0} \left( {1\over \e^{3/2} \sqrt{\gamma}}
{\delta S_{\rm{sub}} \over \delta \Phi (x,\e)}
\right) = - 2 (\phi_{(2)} + \psi_{(2)}) + {2\over 9} \phi_{(0)}^{3}  
\label{gppzovev}
\ee
For the stress energy tensor we obtain
\bea \label{vevtij}
&& \<T_{ij}\>=  g_{(4)ij} + 
{1 \over 8}[\Tr\, {g}_{(2)}^2 - (\Tr\, {g}_{(2)})^2]\, g_{(0) ij} 
- \half ({g}_{(2)}^2)_{ij} 
+ {1 \over 4}\, {g}_{(2) ij}\, \Tr\, {g}_{(2)}
\nonumber \\
&& + [\f_{(0)} (\f_{(2)} -\half \psi_{(2)}) - {1\over 4} (\pa 
\f_{(0)})^{2}] 
g_{(0)ij} 
+ \half \pa_{i}\f_{(0)} \pa_{j}\f_{(0)} 
+ {3 \over 4} ( T^a_{ij} + T^\phi_{ij}) \qquad \qquad
\eea 
where $ T^a_{ij}$ and $ T^\phi_{ij}$ are the stress tensors of the 
conformally invariant trace anomalies. 
Their contribution is scheme dependent. 

Exactly as in our discussion of the stress energy tensor in the 
Coulomb branch 
flow, one can use the asymptotic solution of the bulk field equations 
to compute 
the divergence and trace of the stress energy tensor,
\bea
&&\hspace{-.5cm}
\nabla^i \<T_{ij}\> = -  \<\co\> \nabla_j \f_{(0)} \label{divWGPPZ} \\
&&\hspace{-.5cm}\<T^i_i\>=- \f_{(0)} \< \co \>
+{1 \over 16} (R_{ij} R^{ij} -{1 \over 3} R^2)
-\half [(\nabla \f_{(0)})^2 - {1 \over 6} R \f_{(0)}^2]
\label{trWGPPZ}
\eea
These are the expected Ward identities.
The second term in (\ref{trWGPPZ}), is the holographic gravitational
conformal anomaly (\ref{gravanom}) \cite{HS}
and the last term,
\be \label{matan}
\ca_{scal} = - \half [(\nabla \f_{(0)})^2 - {1 \over 6} R \f_{(0)}^2]
\ee
is a conformal anomaly due to matter \cite{PeSk,KSS,KMM}.
The integrated anomaly should itself be conformal invariant,
and indeed (\ref{matan}) is equal to
the Lagrangian of a conformal scalar. The coefficients
are also the ones dictated by non-renormalization theorems.

In principle $n$-point functions involving $T_{ij}$ plus any set
of additional operators containing $T_{kl}$ and $\co_{\Phi}$ can 
be computed from (\ref{vevtij}). However, the key information
is contained in the generically non-local dependence of $g_{(4)ij}$ 
and $\f_{(2)}$ on the sources. To first order in sources this 
information
can be obtained from the linear fluctuation equations, as was
done in \cite{howtogo}. To higher order, i.e. for 3-point correlators
and beyond, one needs either a fully
non-linear solution of the field equations or a perturbative
formulation via Witten diagrams in which linear fluctuations play
the role of bulk-to-boundary propagators.

Finally we note that supersymmetry requires 
 $\<T_{ij}\>=0$ when sources are set to their values in the domain
wall background. It is this physical requirement that is satisfied
because of the inclusion of the finite 
`counterterm' quartic in $\Phi$.


\section{Vector sector} \label{vecsec}

\subsection{CB Flow} 
\label{gaufxd}

In this section we study gauge fields in the fixed domain-wall 
background
(\ref{CB}). The
physics whose holographic description we seek is that of the 
conserved currents
of a spontaneously broken symmetry and their accompanying Goldstone
bosons. The components of the operator $\Tr X^2$ probe this physics. 
With 
reference to the decomposition of the $20'$ representation of $SO(6)$ 
given
in Sec. \ref{cbdisc}, the $({\bf 1,1})_0$ component has non-vanishing VEV. 
It is
dual to the bulk scalar $\Phi$ discussed in earlier sections on the 
Coulomb
branch flow. The 8 components of $\Tr X^2$ in $({\bf 2,2})_+ \oplus 
({\bf 2,2})_-$ are 
interpolating fields for the Goldstone bosons. The bulk duals of all 
$20'$
components of $\Tr X^2$ can be packaged as a traceless symmetric 
$M_{ij}$ with
the dynamics of a gauged non-linear $\s$-model \cite{bs2}.
The Goldstone components of $M_{ij}$ are related 
to the ``angular'' fields used in our work by (12) of
\cite{bs2}.
  
We thus concentrate on the coset sector of the Coulomb branch. 
In our linearized treatment it is sufficient to use a single vector 
field 
$A_\m$ (and accompanying phase $\a$) to describe any of the 8 vectors 
in this
sector. It is the longitudinal and radial components of $A_\m$ that 
are
relevant to the broken symmetry physics. However it is artificial to
separate them completely from the transverse components and we include
results on these in coset, $SO(2)$ and $SO(4)$ sectors. Final 
transverse
correlators agree with \cite{bs2}. 

The first step in the analysis is the near boundary solution of the 
vector
equations of motion from the action (\ref{action}).
The functions $K$ in the gauge 
kinetic terms all approach
1 near the boundary, i.e. $K=1 + \co(\r)$, as one can verify using 
(\ref{Kfun}) and (\ref{vrho}). The AdS masses of the unbroken vectors 
are 
vanishing $M_{SO(4)}^{2} = M_{SO(2)}^{2} = 0$, while
for the coset vectors we have the asymptotic expansion
\be
M_{X}^{2} =\ell^4 \r^2(1 - {1 \over 3} \ell^2 \r) + \co(\r^4)
\ee

An uncoupled equation for $B_\r$ can be obtained by combining 
(\ref{vecrad})
and (\ref{veclong}). See \cite{Anatomy} for details. Its most 
singular 
solution behaves as $\r^{-1}$ 
near the boundary. From (\ref{veclong})
we learn that the leading behavior of $B_i$ is logarithmic.
We therefore postulate the following expansions
\bea
&&B_\r = {1 \over \r} \left(B_{(0)} 
+ \r (B_{(2)} + \log\r \tilde{B}_{(2)}) + ...\right) \label{rho} \\
&&B_{i} = B_{(0)i} + \log\r \tB_{(0)i} + \r (B_{(2)i} + \log\r 
\tB_{(2)i})
+ ... \label{bi}
\eea
Inserting these expansions in the field equations one 
obtains\footnote{
If one extends the system of equations to include the scalar $\Phi$ 
one
finds that there is a back reaction whose leading singular term is
$\d \Phi=- \sqrt{{3/2}} \r \log^2 \r B_{(0)}^2$. We have shown that 
this
correction may be included in the $\Phi$ counter terms already written
in (\ref{ctgf}). The back reaction then does not affect the 
computation of
current correlators, so it is ignored in order to simplify the present
discussion.}
\bea
&&\tB_{(2)} = -{1 \over 4} \Box B_{(0)}, \quad \tB_{(0)i} = \pa_i 
B_{(0)},
\quad \pa_i B_{(0)i} = - 4 (B_{(2)} + \tB_{(2)}) + {4 \over 3} 
B_{(0)} \ell^2
\nonu 
&&\tB_{(2)i} = \pa_i \tB_{(2)} - {1 \over 4} \pa_j F_{(0)ji}, \quad
\pa_i B_{(2)i} = \Box (B_{(2)} - \tB_{(2)}) - \ell^4 B_{(0)} 
\label{bieqn}
\eea
where $F_{(0)ij}$ is the field strength of $B_{(0)i}$.

We have so far presented the discussion in terms of the gauge
invariant combination $B_\m = A_\m + \pa_\mu \a$\footnote{The scalar 
field $\alpha$ is normalized so as to absorb a factor of the gauge 
coupling which is $g=2$ in our present conventions.}. To derive 
correlation functions of $J^i$ and $O_\a$ 
we need to rewrite the equations in terms of corresponding sources.
We will work in the axial gauge $A_\r=0$. In this gauge one can 
integrate (\ref{rho}) to obtain $\a$,
\be \label{al}
\a(x,\r)=\log \r B_{(0)}(x) + \a_{(0)}(x)+ \r (B_{(2)}(x) - 
\tilde{B}_{(2)}(x)
+ \log \r \tilde{B}_{(2)}(x)) + ...
\ee
$B_{(0)}$ is the source for the dual operator $O_\a$, 
and as we shall shortly see, $\a_{(0)}$, the ``integration constant''
is related to the VEV of  $O_\a$.

{}From (\ref{bi}) and (\ref{al}) we obtain
\be
A_i = A_{(0)i} + \r (A_{(2)i} + \log \r \tilde{A}_{(2)i}) + ...
\ee
where the coefficients satisfy 
\be
\pa_i A_{(2)i} = - \ell^4 B_{(0)}, \quad \tilde{A}_{(2)i} = 
- {1 \over 4} \pa_j F_{(0)ji}
\ee
The expansion of $A_i$ is also valid for both longitudinal and
transverse components.

{}From $\pa_i B_i =\Box \a_{(0)}$ and (\ref{al}) we find
\be \label{rel}
\Box \a_{(0)} = 
- \pa_i A_{(0)i} -4 B_{(2)} + \Box B_{(0)} +{4 \over 3} \ell^2 
B_{(0)} 
\ee

The near-boundary solutions are now inserted in the action cut off at
$\r=\e$, and one finds the following divergences
\be \label{gfxdiv}
S_{\rm{reg}}=\int d^4x \sqrt{g_{(0)}}
\log\e(-{1 \over 8} F_{(0)ij} F_{(0)}^{ij}-  \ell^4 B_{(0)}^2)
\ee
The counterterms are given by
\be
S_{\rm{ct}} = \int d^4 x \sqrt{\g} 
[{m^2(x,\e) \over \log\e}+{1 \over 8} \log\e F_{(0)ij} F_{(0)}^{ij}] 
\ee
We have expressed the gauge invariant quantity $B_{(0)}$ in terms of
the $\s$-model Goldstone field $m(x,\e) =\ell^4 \e \a(x,\e)$ at the 
cutoff. 
This is not
invariant under the residual gauge transformation $\d \a = \d 
\a_{(0)}$,
but this is expected as discussed below.

We can now derive the 1-point function (in the presence of 
sources). 
By varying the renormalized action, with both $\d A_i$ and 
$\d m = \e (\log \e \d B_{(0)} + \d \a_{(0)})$ we obtain
\bea \label{curCBv}
\delta S_{\rm{ren}} &=& -2 \int d^4 x \sqrt{g_{(0)}}\left(
[\delta A_{(0)i} + \pa_i (\d \a_{(0)} + \log \e \d B_{(0)})] 
F_{\r i} \right. \nonu
&& \left. - (\d \a_{(0)} + \log \e \d B_{(0)}) \ell^4
{\a(x,\e) \over \log \e}
+{1 \over 4} \delta A_{(0)i}  \pa_j F_{(0) ji} \log \e \right) \\
&=&-2 \int d^4 x \sqrt{g_{(0)}} \left(\d A_{(0)i}
(B_{(2)i} +\tB_{(2)i} -\pa_i B_{(2)}) - \ell^{4}
\d B_{(0)} \a_{(0)}\right) \nonumber
\eea
where the first line comes from the variation of the regularized bulk 
action and the second from the variation of the counterterms.
The last line was obtained by partial integration of the derivative
of the variation $\d m$ and use of (\ref{bieqn}). Notice 
that all logarithmic divergences cancel. Furthermore, the 
identification
of $A_{(0)i}$ and $B_{(0)}$ as sources and $\a_{(0)}$ as a VEV is now 
clear.
Since the operator dual to $B_{(0)}$ has dimension 2, the source 
should have dimension 2 too, so the dimensionally correct source is 
$a_{(0)}=\ell^2 B_{(0)}$.

Since we are not interested in correlation functions of
the stress energy tensor we can set $g_{(0)ij}=\d_{ij}$.
Going to momentum space, we rewrite (\ref{curCBv}) as
\bea \label{curCBv1}
\delta S_{\rm{ren}} &=& \int d^4 p \left(
\delta A_{(0)i} [                                 
\p_{ij} (-2 B_{(2)j} - \half p^2  A_{(0)j})  
-2 i  {p_i \over p^2} \ell^2 a_{(0)}]  \right. \nonu
&&\left. 
+ \d a_{(0)} \ell^2 [-2 i  {p_i \over p^2} A_{(0)i} 
- {2 \over p^2} (-4 B_{(2)} - {p^2 \over \ell^2} a_{(0)} 
+ {4 \over 3} a_{(0)})]
\right)
\eea
where we have used (\ref{bieqn}) and (\ref{rel}) and the definition
of $a_{(0)}$.
Note that this expression is not invariant under the residual gauge 
transformation $\d A_{(0)i} = i p_i \d \l, \d a_{(0)}=0$, but one 
obtains
$\delta S_{\rm{ren}} = -2 \int d^4 x \ell^2 a_{(0)}$. 
The failure of gauge invariance
is related to the appearance of a Goldstone pole. A similar phenomenon
occurred in the treatment of the gravitational sector for the Coulomb 
branch.
(See (5.7) of \cite{howtogo}. The gauge transformation there is the
linearization of (2.17) for $\d g_{(0)ij}$.).

We can read the transverse, longitudinal and scalar
1-point functions from (\ref{curCBv1}),
\bea 
&&\< J_i \>_{(t)} = \p_{ij} (-2 B_{(2)j} - \half p^2  A_{(0)j}) 
\label{CBtr} \\
&&\< J_i \>_{(l)} = -2 i {p_i \over p^2} \ell^2 a_{(0)} \label{CBlog} 
\\
&&\< O_\a \> = -2 i  {p_i \over p^2} \ell^2 A_{(0)i} 
- {2 \ell^2 \over p^2}  (-4 B_{(2)} - {p^2 \over \ell^2} a_{(0)} 
+ {4 \over 3} a_{(0)}))
\eea
It follows that the 2-point functions of the longitudinal 
currents is zero and the one of the current
with the operator $O_\a$ is trivial up to the expected 0-mass pole,
\be \label{JO}
\<J_i(p) O_\a(-p)\> = 2i \ell^2 {p_i \over p^2}
\ee
This correlator is real in $x$-space, as it should.
The non-trivial 2-point functions of the transverse current 
correlators and the scalar operator will be presented in 
section \ref{cbcorr}.

\subsection{GPPZ Flow} \label{gppzbd}

The physics of this sector is that of the $U(1)_R$ current of an
$\cn =1$ decomposition of $\cn =4$ SYM theory. Current conservation
is spoiled by the mass deformation, so we expect to find that
\be
\nabla^i J_i = q \beta O_{\Psi}
\ee
where $O_{\Psi}$ is a $U(1)_R$ rotation of the mass operator
$O_\Phi$, and $q=-2/3$ is the $R$-charge of $O_{\Psi}$.
These operators are both in the anomaly supermultiplet
of the mass-deformed theory, so we should obtain $\b = -\sqrt{3}$ as
in the gravity sector \cite{howtogo}.

The Lagrangian of this sector was derived from $\cn = 8, D=5$ gauged
supergravity in \cite{Anatomy} and presented 
both in the non-linear St\"{u}ckelberg formalism 
(see (124) of \cite{Anatomy})
and linearized (see eq.(42) of \cite{Anatomy}). 
In our present notation this Lagrangian reads
\be
{ 1 \over \sqrt{G}} \cl =
{ 3 \over 4} F_{\m \n}^2 +{1 \over 2} \left[ (\partial_\m \F)^2
       + 3 \sinh^2\left({2 \F \over \sqrt3}\right)
(A_\m - \half \partial_\m \a)^2\right]
      + V(\F)
 \label{vecact}
\ee
where $K=3$ and the vector mass is defined in (\ref{gppzvecmass}).
Notice that we kept the normalization of \cite{Anatomy} for the 
St\"{u}ckelberg field.
The St\"{u}ckelberg formalism is convenient to solve the equations of
motion, but it somewhat obscures the physics. To bring out the
physics we first linearize near the boundary where $\F \rightarrow 0$,
obtaining
\be
{1 \over \sqrt{G}} \cl =
{3 \over 4} F_{\m \n}^2 +{1 \over 2} \left[(\partial_\m \F)^2
     +4 \F^2 (A_\m - \half \partial_\m \a)^2\right]  +V(\F)
\ee
The change of variables described in Sec. 6.2 of \cite{Anatomy}, 
which at the linearized 
level (and in our present notation) reads
\be
\varphi = \Phi \cos(\alpha) \quad \psi = \Phi \sin(\alpha)
\ee
leads to the quadratic Lagrangian
\be
{ 1 \over \sqrt{G}} \cl =
{ 3 \over 4} F_{\m \n}^2 +{1 \over 2}
[(\partial_\m \varphi + 2 A_\m \psi)^2 + (\partial_\m \psi - 2 A_\mu
\varphi)^2]
\label{cartlag}
\ee
The quadratic potential from (\ref{gppzpot}) must be added. It is
gauge invariant.
 
A rescaling of $A_{\m}$ is needed to bring (\ref{cartlag}) to a  form
in which the $U(1)_R$ generator $T_R$ is conventionally normalized. 
In \cite{Anatomy}  the unusual normalization $tr_{\bf 15}(T_R^2) = 12$
in the adjoint of $SU(4)$ was used, whereas $tr_{\bf 15}(T_R^2) = 4/3$
is conventional in ${\cal N}=1$ field theory. We therefore scale  
$A_\mu \rightarrow A_\mu/3$ giving the Lagrangian
\be
{ 1 \over \sqrt{G}} \cl =
{ 1 \over 12 } F_{\m \n}^2 +{1 \over 2}
[(\partial_\m \varphi + {2\over 3} A_\m \psi)^2 + (\partial_\m \psi - 
{2\over 3} A_\mu \varphi)^2]
\label{cartlagfin}
\ee
With this form of the vector Lagrangian, the source of the standard 
$U(1)_R$ current
is the new $A_\mu$. The field $\psi$ approaches the boundary at the
rate $\r^{{1/2}} \F_B \a$, with $\F_B = \sqrt{3}$ given by
the background value in (\ref{gppzwall}). Thus the source of the 
operator
$\co_\Psi$ is $\sqrt{3}\a$.

We now construct the near-boundary solution of the field equations for
the St\"{u}ckelberg field $B_\m=A_\m -{3 \over 2}\pa_\m \a$. 
The vector equations are
obtained from (\ref{vecact}) after the rescaling. The appropriate 
expansions
are
\bea
B_{i}(x,\rho) &=& B_{(0)i} + \rho B_{(2)i}
+ \rho \log{\rho} \tilde{B}_{(2)i} + \ldots
\\
B_{\rho}(x,\rho) &=& B_{(0)} + \log \rho \tilde{B}_{(0)}
+\rho (B_{(2)} + \log{\rho} \tilde{B}_{(2)}) + \ldots
\label{Brhogppz}
\eea
The transverse components of $B_i$ satisfy an uncoupled equation of
motion while longitudinal and radial components
are coupled. The expansions are substituted and one can solve order
by order in $\r$ to obtain
\bea
&&\tilde{B}_{(2)} = -{1\over 4} (\Box-4) \tilde{B}_{(0)},
\qquad
B_{(2)}= (\half \Box - 1)\tilde{B}_{(0)} -{1\over 4} (\Box-4) B_{(0)},
\nonu
&&\nabla^{i} B_{(0)i} = -4 \tilde{B}_{(0)},
\qquad   \label{Bexp}
\nabla^{i} B_{(2)i} = \Box \tilde{B}_{(0)} - 4 B_{(2)}, \\
&&\tilde{B}_{(2)i} = (1-{1\over 4} \Box) B_{(0)i} . \nonumber
\eea
We will work in the axial gauge $A_\r=0$. In this gauge one can 
integrate
(\ref{Brhogppz}) to obtain $\a(x,\r)$,
\be
\a(x,\r)=-{2 \over 3} \left(
\a_{(0)} - \tilde{B}_{(0)} + \r (B_{(0)} - \half \tilde{B}_{(2)})  + 
\r \log \r \tilde{B}_{(0)} + ... \right)
\ee
where $\a_{(0)}$ is an integration constant. Let us call 
$a_{(0)}=\a_{(0)} - \tilde{B}_{(0)}$. One can now obtain the
expansion of $A_{i}(x,\r)$ but we do not report this formula
here since the explicit expressions for the expansion of 
$B_i(x,\r)$ in (\ref{Bexp}) will be sufficient for our purposes.

Substituting the near-boundary solution in the action (\ref{vecact}) 
we compute the
regulated on-shell action whose counterterms are
\be
S_{\rm{ct}} =
{1\over 6} \int d^4x \sqrt{\gamma}
\log{\e} \left[{1\over 4} \gamma^{ik}\gamma^{jl} F_{ij} F_{kl}
+ {1 \over 2} M^2(\e)  \gamma^{ij} B_i B_j \right].
\ee
As usual $S_{\rm{ren}} = \lim_{\e\rightarrow 0}
[S_{\rm{reg}} + S_{\rm{ct}}]$.

The variation of the renormalized action is
\bea \label{vecvar}
\d S_{{\rm ren}} &=& - {1\over 3} \int \sqrt{g_{(0)}}
\delta B_{(0)i} [B_{(2)i} + \tB_{(2)i} - \pa_i B_{(0)}] \\
&=&  -{1\over 3}\int \sqrt{g_{(0)}}
\left(\delta A_{(0)i} 
[B_{(2)i} + \tB_{(2)i} - \pa_i B_{(0)}]
-6B_{(0)} \delta a_{(0)}\right) \nonumber
\eea
where we have used $B_{(0)i}= A_{(0)i}-{3 \over 2} \pa_i a_{(0)}$ and 
integrated by parts.
Thus we finally obtain
\bea
\<J_{i}\> &=&
-{1\over 3} [ B_{(2)i} + \tilde{B}_{(2)i} - \pa_{i} {B}_{(0)} ] 
\label{jGPPZ} \\
\< O_\a \> &=& 2 B_{(0)} \label{oa}
\eea
They satisfy the Ward identity
\be \label{curWI}
\nabla^{i} \<J_{i}\>  = {4 \over 3}
 B_{(0)} = {2 \over 3} \< O_\a \> =  {2 \over 3} \sqrt{3} 
\< O_\Psi \> = -{2 \over 3} \b \< O_\Psi \>
\ee
where in the first equality we used the (\ref{jGPPZ}) and 
(\ref{Bexp}), in the second (\ref{oa}), in the third the 
relation between the sources of $O_\a$ and of $O_\Psi$ derived
earlier and in the last one the value of the $\b$ function.

We have already derived the Ward identity (\ref{curWI}) using 
the St\"{u}ckelberg gauge invariance  
in section 4.5, but it will be instructive to rederive it 
starting from (\ref{cartlagfin}). The Lagrangian in 
(\ref{cartlagfin}) is uniquely fixed by gauge invariance 
and the requirement that in the UV limit the correlators
approach that of $\cn=4$ SYM with standard normalizations.
The charges of $\f$ and $\psi$ under $U(1)_R$ are opposite to 
that of the operators they couple to. The Lagrangian (\ref{cartlag})
is invariant under the following gauge transformations
\be
\d \varphi = -{2 \over 3} \l \psi, \qquad
\d \psi = {2 \over 3} \l \phi, \qquad
\d A_\mu = \pa_\mu \l
\ee
Evaluating on the background we get that 
the sources transform as
\be
\d \varphi_{(0)}=0, \qquad 
\d \psi_{(0)} = {2 \over 3} \l \Phi_B,  \qquad \d A_\mu = \pa_\mu \l
\ee
where $\varphi_{(0)}$ and $\psi_{(0)}$ are 
the sources for $O_\Phi$ and $O_{\Psi}$, respectively, and $\Phi_B$ 
is the background value of $\Phi$. This then leads directly to
the Ward identity
\be
\nabla^{i} \<J_{i}\> 
= {2 \over 3} \Phi_B \<O_\Psi \> = - {2 \over 3} \b \<O_\Psi \> 
\ee
This derivation should be contrasted with the one 
of the relation $\<T^i_i\> = \beta \<O_\Phi \>$ presented in 
\cite{howtogo}.
There Weyl invariance implied
$\<T^i_i\> = - \phi_{(0)} \<O_\Phi \> + \ca$, which upon 
evaluation on the background solution lead to the desired 
$\b$-function relation.


\section{Linearized analysis around domain-walls}
\setcounter{equation}{0}

As remarked several times, the near boundary analysis does not fix 
certain asymptotic coefficients, 
associated with operator vevs, in terms of the 
corresponding sources. One needs a solution of 
the field equations which is valid beyond the asymptotic region 
of small $\r$. For the purpose of computing 
2-point functions a linearized solution around a given background 
is enough. 
We have already used this strategy in the gravitational sector in
\cite{howtogo},  and we now 
extend it to vector fields. We will thus 
recall from \cite{Anatomy} the relevant linear equations and their 
solutions.

As shown in \cite{Anatomy}, for supersymmetric flows, 
transverse modes of the metric and vector fields 
\footnote{
Let us mention that by
rescaling the vectors so that their kinetic term is canonically 
normalized,
one obtains that all transverse vector fluctuations 
have the common mass \cite{Anatomy} 
$$
M_{eff}^{2} =  \half {\pa_r^2 K \over K} 
+ A' {\pa_r K\over K} - {1\over 4} 
\left( {\pa_r K \over K}\right)^{2} + {M^{2}\over K} = - 2 \pa_r^2 A.
$$
We have checked that this relation is satisfied by 
the graviphoton in the GPPZ flow and by any vector field in all known
supersymmetric CB flows with one active scalar.}
can be expressed in terms of an auxiliary ``massless scalar field'' 
$f(r)$.
For transverse traceless tensor fluctuations one has
\be
h_{ij}^{TT}(r,x) = e_{ij}(p)e^{ip\cdot x}f_p(r)
\ee
while for transverse vectors one finds
\be
a_{i}^{T}(r,x) = v_{i}(p) e^{ip\cdot x}K^{-1/2} e^{2A}{df_p \over dr}
\ee
Longitudinal and radial modes are less universal and must be studied 
on a case by case basis.

\subsection{CB flow} 
\label{CBcor}

With $v$ as radial variable, the ``massless'' scalar field equation 
has solution
\be
f_{p}(v) = v^{a} F(a,a;2a+2;v) 
\ee
where
\be
a = -{\half} +\half \sqrt{ 1 + {p^{2} \over \ell^{2}}}
\label{defap}.
\ee

The transverse vector fluctuations thus read
\be \label{trfl}
b^{(t)}(v) = 
v^{\l+a-1} F(a,a+1;2a+2;v)
\ee
where $\lambda=1$ for $SO(4)$, $\lambda=2$ for $SO(2)$ and 
$\lambda=3/2$ for the coset. 
The `unbroken' $SO(4)\times SO(2)$ vectors have no physical 
longitudinal 
fluctuations. The longitudinal components 
of the coset vectors can be expressed in terms of the radial component
$B_{r}$. In turn, $B_{r}$ can be rewritten as 
\be
B_{r}(r,x) = e^{-4A} M^{-2} C_p(r)e^{ip\cdot x}
\label{defbc}
\ee
When expressed in terms of the $v$ variable, the scalar function 
$C_p(v)$ satisfies
\be
v^{2}(1-v) C_p'' + v(1-v) C_p' - {1\over 4} [ {p^{2}\over \ell^{2}} + 
(1-v)] C_p = 0
\ee 
whose solution is the hypergeometric function
\be \label{cCB}
C_p(v) = v^{a+1/2} F(a, a+1, 2a+2; v) 
\ee
with $a$ defined in (\ref{defap}). Up to (irrelevant) overall 
constant,
\be \label{Bv}
C_p(v)=v(1-v) B_v
\ee
The coordinate transformation (\ref{vrho}) may be used to obtain 
$B_{\r}$. 

\subsection{GPPZ flow} \label{gppzfl}

When expressed in terms of the radial variable $u$, the ``massless'' 
scalar 
field equation admits the solution
\be
f_{p}(u) = (1-u)^2 F(2+i{p\over 2}, 2-i{p\over 2}; 2; u) \; .
\ee
For the transverse components of the $U(1)_{R}$ graviphoton one has 
\be \label{gppztr}
b^{(t)}(u) = u (1-u) F(2+i{p\over 2}, 2-i{p\over 2}; 3; u)
\ee
after dropping an irrelevant $p$-dependent factor.

For the longitudinal and radial components one finds
\be \label{CGPPZ}
C_{p}(u) = -8 B_\r =
u F\left({3\over 2} + \half q, {3\over 2} - \half q; 3; u\right)
\ee
where
\be
q = \sqrt{1 - p^{2}}
\ee


\section{Correlation functions of vector currents}
\setcounter{equation}{0}

In this section we will obtain results on the 2-point correlation
function of various vector currents and their scalar partners in the
$CB$ and $GPPZ$ flows. We will study this system in the fixed 
background
approximation in which the back reaction on the background metric and
active scalar is neglected. It was argued earlier
that the fixed background approximation is sufficient to compute all
correlators in this sector except those involving $T_{ij}$.
 
It is actually easy to see that the 2-point function $\<T_{ij}(p) 
J_k(-p)\>=0$,
so that nothing is missed in the fixed background treatment.
On the gravity side this correlator could only come from bilinear 
terms
$h_{ij} A_k$ in $S_{{\rm ren}}$, and such terms are clearly absent. 
For the
field theory of Coulomb branch flow, $\<T_{ij}(p) J_k(-p)\>$ must be 
conserved in all 3 indices,
and there is no tensor structure with this property. For the GPPZ flow
the tensor structure $\<T_{ij}(p)J_k(-p)\>=\pi_{ij} p_k A(p^2)$ is 
allowed.
However we trace on $ij$ and contract with $p_k$ and use the operator
equations of the theory we find that $A(p^2)$ is proportional to the
2-point function of a scalar and pseudoscalar operator which vanishes
by parity conservation.

\subsection{CB flow} 
\label{cbcorr}

For the CB flow the physics of the  `unbroken' and `coset' 
currents, is different. The distinction between sectors is almost 
irrelevant for transverse components. Since only coset vectors have
longitudinal components, we can treat all sectors together
with a small elaboration of notation.

The first step is to perform the asymptotic expansion of the 
hypergeometric 
functions that appear in the solutions of the linearized field 
equations. From (\ref{trfl}) one finds that the transverse vector 
fluctuation
behaves as 
\be
A_i = A_{(0)i}\left(1 - \rho \ell^{2} [\lambda -1 
+ {p^2 \over 4 \ell^2} -
{p^{2}\over 2 \ell^2 } (\psi(a+1) - \psi(1) + \half \log \rho) ]
+ \ldots \right)
\ee
where $\lambda=3/2$ for the coset vectors,  while for the `unbroken' 
vectors $\lambda=0$ for $SO(2)$ and $\lambda=1$ for $SO(4)$.
For the radial component in the coset sector, one can use (\ref{cCB})
to obtain
\be \label{bp}
B_{(2)}=a_{(0)} \left(-{1  \over 6} 
- {p^2 \over 4 \ell^2}+{p^2 \over 2 \ell^2} (\psi(a+1)-\psi(1)) 
\right).
\ee
This information is sufficient to fix $B_{(2)i}$ and $B_{(0)}$ which 
were
the two unknown coefficients in (\ref{curCBv1}). This variational 
expression
may now be ``integrated'' to obtain the bilinear fluctuation action
\bea \label{flact}
S_{{\rm ren}} &=& {N^2 \over 2 \p^2}
\int d^{4}p \left( A_{(0)i}(p) \p_{ij} A_{(0)i}(-p)
\left(-\ell^{2} (\lambda-1) + {p^{2}\over 2}[\psi(a+1) - \psi(1)] 
\right) 
\nonumber \right.\\
&& \hspace{-2cm}\left.
- \delta_{X} 2i \ell^{2} a_{(0)}(p) {p^{i}\over p^{2}} A_{(0)i} (-p)
+[-{2 \ell^{2} \over p^2} 
+ 2 (\psi(a+1) - \psi(1))] a_{(0)}(p) a_{(0)}(-p) \right)
\label{actflu}
\eea
where we have reinstated the overall factor of $N^2/2 \p^2$, and 
where 
$\delta_{X} = 0$ for the `unbroken' $SO(4)\times SO(2)$ vectors 
reflecting the absence of physical longitudinal modes, while 
$\delta_{X} = 1$ 
for the coset vectors.

The negative of the second variation of $S_{{\rm ren}}$ 
wrt to sources gives the 
the desired 2-point functions. 
For the $SO(4)\times SO(2)$ longitudinal modes one obviously gets 
\be
\langle J^{i}(p) J^{j}(-p) \rangle_{(l)} = 0
\ee
since there is no mixing with scalars. 
For the 2-point function of current and the scalar operator, we can
repeat (\ref{JO}), slightly generalized to read 
\be \label{JO2}
\<J_i(p) O_\a(-p)\> = {N^2 \over 2 \p^2} \d_X 2i \ell^2 {p_i \over 
p^2}
\ee

The 2-point function of the operators dual to the coset scalars 
is obtained from (\ref{actflu}). It reads
\be
\langle O_\a(p) O_\a(-p) \rangle= {N^2 \over 2 \p^2}
\left({4 \ell^{2} \over p^2} 
- 4 (\psi(a+1) - \psi(1))\right)
\ee
This has the expected massless pole with a residue proportional to 
$\ell^2$ and also the correct UV behavior.

The transverse current correlator is
\be
\langle J_{i}(p) J_{j}(q) \rangle_{(t)} = {N^2 \over 2 \p^2} \p_{ij}
\left(-2 \ell^{2} (\lambda-1) + p^{2}[\psi(a+1) - \psi(1)] \right).
\ee
It has the correct scaling in the UV but it also
has a massless pole except for $\lambda=1$, {\it i.e.} for the $SO(4)$
currents. The result agrees with \cite{bs2}.
Although the supergravity solution ('continuous' disk distribution of
D3-branes) seems to preserve an $SO(4)\times SO(2)$ subgroup of 
$SO(6)$, no discrete disk distribution can preserve more
than $SO(4)$. The field-theory counterpart of this phenomenon is the 
absence
of any locus in the Coulomb branch of $\cn = 4$ SYM with any finite
number of colors $N$ on which $SO(4)\times SO(2)$ is preserved. This 
is
discussed in the Appendix. We thus interpret   
the massless poles as Goldstone poles in both the coset and $SO(2)$ 
sectors.
The latter indicates that supergravity implicitly agrees with field 
theory,
in accord with the discussion in the Addendum to \cite{bs2}.

\subsection{GPPZ flow}

In this section we combine results of near boundary analysis from
section \ref{gppzbd} with the fluctuations given in section 
\ref{gppzfl} 
to obtain
correlation functions of the $U(1)_{R}$ current and the scalar 
operator
$O_\a$.

It is again useful to `integrate' the first order variation 
$\delta S_{{\rm ren}}$ 
in (\ref{vecvar}) to obtain a quadratic action from which the 
2-point correlators may be immediately read. We first
determine the non-local relation between $B_{(2)i} + 
\tilde{B}_{(2)i}$ 
and $A_{(0)i}$ from the asymptotics of the hypergeometric functions 
given in section \ref{gppzfl}
as $u \rightarrow 1$. For the transverse components
(\ref{gppztr}) gives
\be
B_i = A_{(0)i} ( 1 - \rho [(1+{p^2 \over 4}) [(\bar{K}-\half)
- \log\rho)] + 1] + ...)
\ee
where $\bar{K} = \psi(3) + \psi(1) - \psi(2+ip/2) - \psi(2-ip/2)$.
One then finds
\be
\tilde{B}_{(2)i} = A_{(0)i} (1+{p^2 \over 4})
 \qquad
B_{(2)i} = - A_{(0)i} [(1+{p^2 \over 4})(\bar{K}-\half) +1]
\ee
Similarly, for the radial component, (\ref{CGPPZ}) gives
\be
B_{(0)} = - \tilde{B}_{(0)} \bar{J},
 \ee
where $\bar{J}=2 \psi(1) - \psi(3/2+q/2) - \psi(3/2-q/2)$
and from the near-boundary analysis we know that
\be
\nabla^{i} B_{(0)i} = - 4 \tilde{B}_{(0)}
\ee

The renormalized action to quadratic order in the sources thus reads
(in momentum space) 
\bea
S_{{\rm ren}} &=& {N^2 \over 12 \p^2} 
\int d^{4}p \left(  A_{(0)i} \pi^{{ij}} A_{(0)j}
[-(1+{p^2 \over 4}) (\bar{K}-\half)+ {p^2 \over 4}] \right. \nonu
&&\left.-{1 \over 2 p^2} (p_i A^{i}_{(0)}+ {2 \over 3}
i p^2 a_{(0)})^2 \bar{J} \right)
\eea
The fact that the the longitudinal part of the $A_{(0)i}$ and the 
$a_{(0)}$ appear as a total square is a consequence of the beta 
function operator relation (\ref{beta}) between the divergence of 
$J_i$ and
$O_\a$. We have seen an analogous phenomenon in the graviton-scalar
section in \cite{howtogo}.

The transverse 2-point function reads
\be
\< J_i(p) J_j(-p)\>_{(t)} = {N^2 \over 6 \p^2} 
\pi_{ij} \left((1+{p^2 \over 4}) (\bar{K} -\half)-{p^2 \over 4} 
\right)
\ee
It has poles at $p^2 = - 4 (n + 2)^2$ (we are  
using Euclidean signature) with 
$n=0,1,...$, as expected, but also a disturbing  
massless pole whose
residue is $-(N^2/6\p^2)
(\bar{K}(p{=}0)-1/2) = (N^2/6\p^2)$.
Happily, as we now show, the longitudinal 2-point function
also contains a massless pole and the two contributions
cancel each other!

The remaining correlators are 
\bea
\< J_i(p) J_j(-p) \>_{(l)} &=& {N^2 \over 12 \p^2} {p_ip_j \over 
p^2}\bar{J} \\
\< J_i(p) O_{\a}(-p) \> &=& {N^2 \over 18 \p^2} i p_i  \bar{J}
\label{JiOa} \\
\< O_\a(p) O_\a(-p) \> &=& - {N^2 \over 27 \p^2} p^2 \bar{J}
\label{OaOa}
\eea
The residue of the zero mass pole in $\< J_i(p) J_j(-p) \>_{(l)}$
is $N^2/(6 \p^2)$,
and indeed the zero mass poles cancel. 

These correlation functions are consistent with the operator 
relation
\be \label{beta}
\nabla^i J_i = {2 \over 3} O_\a = -{2 \over 3} \b O_\Psi
\ee 
where $\b=-\sqrt{3}$ is the same beta function found in 
the graviton-scalar sector in  \cite{howtogo}.
The current Ward identity (\ref{curWI}) is related by supersymmetry 
to the trace Ward identity, $T^i_i = \b O_\F$.



\section*{Acknowledgments}
We would like to thank M. Berg for a discussion of the results 
presented in \cite{berg} prior
to publication and for finding some typographical errors
in an early draft of this paper.
Part of the present work was done during the summer workshop 
``Physics in the Pyrenees: 
Strings, Branes and Field Theory'' held in Benasque, Spain. 
We thank 
the local organizers and other participants for the stimulating 
atmosphere there. The collaboration of M.B. and D.Z.F. was 
facilitated 
by the INFN-MIT `Bruno Rossi' exchange program.
Much of this work was carried out while M.B. was visiting 
D.A.M.T.P. at Cambridge University. M.B. would like to thank the 
colleagues at D.A.M.T.P. and in particular Michael Green for 
their kind hospitality and for financial support through a PPARC 
grant.
The work of M.B. was supported in part by the EEC contracts
HPRN-CT-2000-00122 and HPRN-CT-2000-00148, and in part by the
INTAS contract 99-1-590. DZF's research is supported by US NSF grant
PHY-0096515. KS is supported in part by the National Science 
Foundation grant PHY-9802484.


\appendix

\section{Appendix}

In this Appendix we use simple group-theoretic methods to prove a
lemma concerning the Coulomb branch of ${\cal N}=4$ SYM
theory with any finite dimensional gauge group G.

{\it Lemma}: The Coulomb branch does not contain configurations where 
the
$SO(6)$ flavor symmetry is broken to a non-trivial direct product
$SO(n)\times SO(6-n)$ with $2 \le n \le 4$.

The proof assumes the validity of a classical description of the 
moduli
space in terms of the VEV of the elementary scalars  $X^i = X^i_aT_a$.
The VEV  $\langle X^i \rangle$ is in the Cartan
subalgebra $\langle X^i \rangle = v^i_IH_I$.

Without loss of generality we can consider the subgroup $SO(2) \subset
SO(n)$ which rotates in the $i=1,2$ plane. This acts as a complex 
rotation
of the sum $Z=X^1+iX^2$, and preserving $SO(n)$ symmetry requires
\be
e^{i\varphi}Z = g Z g^{-1}
\ee
where $g$ is a gauge transformation depending on $\varphi$. For 
infinitesimal
$\d \varphi$ this requires a Lie algebra element $A$ such that
\be
\d \varphi Z = [A,Z]
\ee
We can  write $A = A^{(+)} +A^{(-)} + A^{(H)}$, where $A^{(+)}$ 
belongs
to the positive root space, $A^{(-)}$ is its adjoint in the negative
root space and $ A^{(H)}$ is in the Cartan subalgebra. Thus
\be
\d \varphi Z = [A,Z] = B^{(+)} - B^{(-)}
\ee
where $B^{(\pm)}$ are in the positive(negative) root spaces and are 
linear
in $Z$. However the last equation is impossible to satisfy unless 
$Z=0$.

Thus for any $SO(n)$-invariant vacuum, the non-vanishing components of
$X^i$ are restricted to $i=n+1,..6$. the same argument may now be 
applied to these components which also must vanish.

We would now like to rephrase the above discussion in terms of gauge 
invariant composite operators. In this approach the question is whether 
it is possible to pick VEV's of the elementary fields $X^i$ such that 
the only composite operators that acquire a VEV are singlets under 
$SO(n) \times SO(6-n)$. The composite operators that admit a supergravity description 
are the so-called chiral primary operators (CPO). They are of the 
form $\Tr(X^{k})$, \ie $k$-fold symmetric traceless products 
of the fundamental of $SO(6)$, that belong to the representation with 
Dynkin labels $[0,k,0]$. Under the decomposition of $SO(6)$ 
R-symmetry into $SO(n) \times SO(6-n)$ the operators 
with $k$ even contain singlets under $SO(n) \times SO(6-n)$
but the ones with $k$ odd contain singlets of each factor only but not of 
the product. So in order to preserve $SO(n) \times SO(6-n)$ 
the VEV's of $X^i$s should be such that (i) for even $k$ 
only the singlet operator gets a VEV, (ii) 
the VEV of all CPOs with odd $k$ are zero.
We will now show that it is not possible to pick VEV's such that these
conditions are satisfied. 

Let us consider the case of a $U(N)$ gauge group. 
We will comment on the generalization to $SU(N)$ afterwards. 
The lowest CPO's in the ${\bf 20'}$ of $SO(6)$ 
contain one singlet of $SO(n) \times SO(6-n)$, say 
$\Tr ((6-n) [X_{1}^{2} + \ldots X_{n}^{2}]- 
n [X_{n+1}^{2} + \ldots X_{6}^{2}])$. In order for the other 
non-singlet components to vanish one has to choose the scalar VEV's so
that the six $N$-component vectors $X_{i}$ be orthogonal to one 
another  and normalized so that $|X_{1}|^{2} = \ldots = |X_{n}|^{2} \neq 
|X_{n+1}|^{2} = \ldots = |X_{6}|^{2}$. 
Let us now turn to the cubic Casimir $\Tr (X_{i} X_{j} X_{k} - 
\ldots)$ in the ${\bf 50}$ of $SO(6)$. 
It does not contain singlets of the product subgroup
$SO(n)\times SO(6-n)$, so all VEV's of 
the CPO in the ${\bf 50}$ should vanish. The only way to achieve this 
is to take $X_{i}$ along the generator $H_{i}$ 
and $X_{1}= \ldots = X_{n} \neq X_{n+1} = \ldots = X_{6}$. The next 
CPO in the ${\bf 105}$ of $SO(6)$ is associated to the quartic 
Casimir $\Tr (X_{i} X_{j} X_{k} X_{l}- \ldots )$.  
Let us consider the complex  field  $Z = X^{1} + i X^{6}$.  
It is easy to check that the operator $\Tr [Z^{4}]$ is not a singlet of 
the product subgroup, and one cannot further restrict the VEVs of $X^i$
(without setting all of them to zero) such that it vanishes.
We thus conclude that R-symmetry cannot be broken to 
$SO(n) \times SO(6-n)$, and 
at least one of the two factors, say $SO(6-n)$, should also be broken.
In the case of $SU(N)$ the only additional complication is that the
tracelessness condition $\Tr H_i=0$ should be taken into account.
This imposes a further constraint among the VEV's that is however 
compatible with the previous choices.

\end{document}